\documentclass[journal,comsoc]{IEEEtran}
\usepackage[T1]{fontenc}% optional T1 font encoding

\usepackage{cite}
\ifCLASSINFOpdf
   \usepackage[pdftex]{graphicx}
\else
 \usepackage[dvips]{graphicx}
\fi
\usepackage{amsmath}
\usepackage[cmintegrals]{newtxmath}
\usepackage{algorithm}
\usepackage{algorithmic}
\ifCLASSOPTIONcompsoc
  \usepackage[caption=false,font=normalsize,labelfont=sf,textfont=sf]{subfig}
\else
  \usepackage[caption=false,font=footnotesize]{subfig}
\fi
\usepackage{url}
\usepackage{threeparttable}

\begin{document}
\title{Modeling e-Learners' Cognitive and Metacognitive Strategy in Comparative Question Solving}

\author{Feng~Tian,~\IEEEmembership{Member,~IEEE,} Jia~Yue, Kuo-ming~Chao,~\IEEEmembership{Member,~IEEE,} Buyue~Qian, Nazaraf~Shah, Longzhuang~Li, Haiping~Zhu,~\IEEEmembership{Member,~IEEE,} Yan~Chen, Bin~Zeng, Qinghua~Zheng,~\IEEEmembership{Member,~IEEE}% <-this % stops a space
\thanks{F. Tian, J. Yue, B. Qian, H. Zhu, Y. Chen, B. Zeng and Q. Zheng are with the MOEKLINNS Lab, School of Electronic and Information Engineering, Xi'an Jiaotong University, Xi'an 710049, P. R. China. (e-mail: fengtian@mail.xjtu.edu.cn; yuejiayyzz@gmail.com; qianbuyue@mail.xjtu.edu.cn; zhuhaiping@mail.xjtu.edu.cn; chenyan@mail.xjtu.edu.cn; zengbin23@126.com; qhzheng@mail.xjtu.edu.cn).}% <-this % stops a space
\thanks{K.M. Chao and N. Shah are with the Faculty of Engineering and Computing, Coventry University, Priory Street, Coventry CV1 5FB, United Kingdom. (e-mail: csx240@coventry.ac.uk; aa0699@coventry.ac.uk)}% <-this % stops a space
\thanks{L. Longzhuang is with the Department Of Computing Sciences, Texas A\&M University-Corpus Christi, TX, USA. (e-mail: longzhuang.li@tamucc.edu)}}

\maketitle

\begin{abstract}
Cognitive and metacognitive strategy had demonstrated a significant role in self-regulated learning (SRL), and an appropriate use of strategies is beneficial to effective learning or question-solving tasks during a human-computer interaction process. This paper proposes a novel method combining Knowledge Map (KM) based data mining technique with Thinking Map (TM) to detect learner's cognitive and metacognitive strategy in the question-solving scenario. In particular, a graph-based mining algorithm is designed to facilitate our proposed method, which can automatically map cognitive strategy to metacognitive strategy with raising abstraction level, and make the cognitive and metacognitive process viewable, which acts like a reverse engineering engine to explain how a learner thinks when solving a question. Additionally, we develop an online learning environment system for participants to learn and record their behaviors. To corroborate the effectiveness of our approach and algorithm, we conduct experiments recruiting 173  postgraduate and undergraduate students, and they were asked to complete a question-solving task, such as ``What are similarities and differences between array and pointer?'' from ``The C Programming Language'' course and ``What are similarities and differences between packet switching and circuit switching?'' from ``Computer Network Principle'' course. The mined strategies patterns results are encouraging and supported well our proposed method.
\end{abstract}


% \fi
\IEEEpeerreviewmaketitle

\section{Introduction}
\IEEEPARstart{I}{n} recent years, cognition, metacognition and their strategies have attracted a great deal of research interests from academia and industry \cite{kim2016applying,2le2018cognitive,5winne2010bootstrapping}. Cognition is about a thinking process, and metacognition is defined as thoughts about thoughts or cognition about cognitions \cite{41Flavell1979Metacognition}. The lack of e-learners' cognitive and metacognitive strategy monitoring or without self-monitoring/self-assessment often leads to lower learning efficiency \cite{1Simon2004Learning,3Metcalfe2013Metacognition}. Hence, it is crucial to detect and model cognitive/metacognitive strategies, in order to make better human-computer interaction, such as providing a means for intelligent tutors to assess learners' current knowledge and skill levels, or helping students improve learning experience and effectiveness. Traditionally, self-report questionnaires \cite{13pintrich1993reliability,14weinstein1987lassi} and process tracing methodologies \cite{17hadwin2007examining,19perry2006learning} are adopted to analyze the learners' learning behaviors or cognition. More recently, researchers have been developing data-mining techniques, including hidden Markov models \cite{21kinnebrew2011comparative}, sequential pattern mining approaches \cite{24biswas2013analyzing,25kinnebrew2014analyzing}, coherence analysis \cite{26segedy2015using}, clustering \cite{27perera2009clustering,28gavsevic2017detecting,29fern2010mining} and fuzzy decision trees method \cite{30crockett2017predicting}. 

However, in most extant research aforementioned modeling cognitive and metacognitive strategy \cite{29tian2007personalized,9josyula2009modeling}, and cognitive activities (goal-setting \& planning, knowledge construction, monitoring and help-seeking) \cite{24biswas2013analyzing,23segedy2011modeling} were usually modeled in a general cognitive strategy level along with its procedural knowledge (e.g., cognitive activity-based sequence pattern), while the metacognitive strategy sits at the same level with cognitive activities  absence of hierarchical model hinders manifestation of their relationships. We agree with the idea, proposed by Kinnebrew et al. \cite{22kinnebrew2013contextualized}, that raising the level of abstraction with context summarization from raw log events to a canonical set of distinct actions is a vital first step for effective analysis. Even though the issue with hierarchy has been resolved, there still exist three main obstacles. Firstly, cognitive strategies can be developed in a generic way to serve different domains or specific to a domain, but the current approaches of modeling specific cognitive strategy seldom consider the domain knowledge presentation or construction for question solving when they apply general cognitive strategy. Secondly, metacognitive strategies are in form of complex constructs, not directly observable and problematic to extract. Current research mainly focuses on human expert's interpretation and manual mapping between cognitive strategies and metacognitive strategies \cite{24biswas2013analyzing,23segedy2011modeling,10biswas2010measuring,20kinnebrew2013investigating}. There is a lack of systematic enabling method that automates mapping of cognitive strategies to these metacognitive strategies by considering knowledge constructions in the context of the selected subject, scope, and sequence (cognition strategy knowledge and domain knowledge itself), when abstracting learner cognitive strategies from a raw log. Thirdly, there lacks a visual language to help portray cognitive and metacognition process and their characteristics, especially for learning approaches.

Aimed at addressing these three problems, an approach to combine Thinking Maps (TM)\cite{33Hyerle2014Thinking} with Knowledge Maps (KM) \cite{34Liu2012Topological,35O2002Knowledge} is proposed to explore the structure and process of e-learner cognitive strategy by applying TM-based high-level abstraction in the context of KM. Generally, TM is a common visual language for learning communities to describe and visualize the thinking process, and a KM is a directed graph as well as a semantic network composed of nodes and edges. Each node is a knowledge unit, and each edge represents a semantic relation between two knowledge units. More detail on this will be discussed in Section III-C.

Combining TM and KM may throw a light on detecting or recognizing individual cognitive and metacognitive strategy simultaneously, especially in the scenarios of question solving. Inspired by this idea and based on our team's prior works \cite{29tian2007personalized,34Liu2012Topological,36Zheng2011Yotta,37Zeng2014A,38Huang2014Generating}, we introduced the theory of TM to detect and model cognitive and metacognitive strategies with learners' question solving behaviors acquired from our KM-based e-learning system which includes the proposed mining methods for cognitive and metacognitive strategies. This paper mainly focuses on how learners solve questions by using their cognitive and metacognitive strategies, so we eliminate factors such as ambiguity of natural language, and complexity of Natural Language Processing (NLP) to obtain the effectiveness of the proposed methods and algorithms. In the experiments, simple questions, like ``What are similarities and differences between array and pointer?'' and ``What are similarities and differences between packet switching and circuit switching?'' are used to test learners and the system.

Below, we summarize our major contributions in this paper.
\begin{enumerate}
	\item Propose a novel method that introduces a theory of thinking maps into the KM-based data mining framework to detect and model cognitive and metacognitive strategies in  scenario of question solving. In which, a Knowledge Map is considered as a domain-specific semantic knowledge structure that is not only a kind of knowledge construction used for learning, but also a knowledge base which an e-learner selects the learning scope by using his/her cognitive strategies to solve questions. Thinking Maps is a set of generic cognition strategies with specific knowledge units and it acts as a bridge for mapping cognition strategies to metacognitive strategy that can be automated.
	\item Provide a graph-shape description to different cognitive strategies based on TM (cognitive planning with specific knowledge units) as well as a sequence-style metacognitive strategy (procedural knowledge of individual cognitive strategy) based on mined results. The TM theory provides a visual language to help reveal e-learners' question-solving behaviors and process of cognitive and metacognitive. Their characteristics are not only exhibited in a general-level but also in a domain-specific-level, simultaneously. The mined results are comprehensive so that they can help other students with learning difficulties to compare and evaluate their cognitive and metacognitive strategies with others.
	\item Coin the corresponding definitions for designed data mining algorithms. We developed an online learning environment for learners and logged their learning behaviors when they carry out assignments task or solving questions. The experimental results are quite encouraging, as they demonstrate the effectiveness of our proposed frameworks and corresponding methods.
\end{enumerate}

The rest of the paper is organized as follows. Section II introduces some of related terminology of cognitive psychology, and then reviews the state-of-the-arts methods of detecting cognitive and metacognitive strategy and their limitations, and then proposes our method. Section III is devoted to explaining the motivation of this paper with the definition of 11 key concepts through a representative example, presenting our algorithm for modeling and detecting cognitive and metacognitive strategy. Section IV describes the experimental setups and summarizes the important results. Section V presents the conclusion, discuss the limitation and future works.

\section{Background and State of the Art Review}
\label{Background and State of the Art Review}

First, we will briefly introduce the basic notions, including cognition, metacognition, cognitive strategy and metacognitive strategy, general cognitive strategy and domain-specific cognitive strategy, as well as their comparison. After that, related works and detecting methods are reviewed in Section II-D.

\subsection{Cognition and Cognitive Strategy}

Cognition is a ``mental process of acquiring knowledge and understanding through thought, experience, and the senses'' \cite{39ocford}. The thinking processes analyzed from different perspectives within different contexts, notably discussed in the fields of psychology, education and philosophy, etc. The term ``cognitive strategies'' in its simplest form is the use of the mind (cognition) to solve a question or complete a task. Pressley and Woloshyn \cite{40Pressley1995Cognitive} hold the view that cognitive strategies can be general or specific. General cognitive strategies can be applied across many different disciplines and situations, such as summarization or setting goals for what to accomplish, whereas specific cognitive strategies tend to be narrow, and they are specific to a task or a domain, such as drawing a picture to help one see how to tackle a physics problem. A specific cognitive strategy should also include the representation or construction of domain knowledge.

\subsection{Metacognition}

In the 1970s, Flavell et al. \cite{41Flavell1979Metacognition} coined the definition of metacognition as ``cognition about cognitive phenomena,'' or more simply ``thinking about thinking''. Metacognition refers to higher order thinking that involves active control over the cognitive processes involved in learning. Metacognition has two constituent parts: knowledge about cognition and cognitive regulation \cite{42Cross1988Developmental,43Paris1990Promoting,44Schraw1995Metacognitive}. Knowledge of cognition includes declarative cognitive knowledge and procedural cognitive knowledge \cite{45Schraw2006Promoting,46Deanna2004Metacognition}.

Procedural knowledge involves awareness and management of cognition, including knowledge about strategies \cite{45Schraw2006Promoting}. Cognitive regulation concerned individual ability of planning, monitoring or regulating, and evaluating is a component of metacognition. Especially, the ability of planning means that learner can identify and select appropriate strategies and allocate resources for learning. This feature enables procedural knowledge to interact with strategy knowledge to apply different types of knowledge according to findings \cite{41Flavell1979Metacognition} (e.g., according to the current outcome, it is better to use strategy A over strategy B to carry out task X). Researchers have found that problem-solving activities require self-regulated learning to be aware of the problem itself and to manage one's cognitive processes, and they suggested that metacognition can guide goal-directed thinking when people are learning about or solving their problems \cite{kim2016applying}.

\subsection{Comparison of Cognitive Strategy and Metacognitive Strategy}

Cognitive strategies are the basic mental abilities we use to think, study, and learn (e.g., making associations between or comparing different pieces of information), which provide a structure for learning how to solve non-intuitive problems. The strategies help an individual achieve a goal (e.g., solving a math question) by analyzing and evaluating the required information individually. In contrast, metacognitive strategies are used to ensure that an overarching learning goal can be achieved. Examples \cite{47Montague1992The} of metacognitive activities include planning of how to approach a learning task, using appropriate skills and strategies to solve a problem, utilizing self-assessing and self-correcting in response to the self-regulation, evaluating progress toward the completion of a task, and becoming aware of distracting stimuli etc. The required procedural knowledge of cognition to fulfill these abilities, however, is more complex than a sequence-shape structure.

One of cognitive strategies, self-monitoring, can be used to monitor metacognitive strategies. For example, one of the major functions of self-regulation is to check themselves if they have solved questions. However, metacognitive strategies with complex constructs are not easy to extract due to their implicit semantics. This leads to that the existing research cannot automate the process of mapping between cognitive strategies and metacognitive strategies \cite{20kinnebrew2013investigating,23segedy2011modeling}, but manually.

\begin{figure*}[!t]
	\centering
	\includegraphics[width=6in]{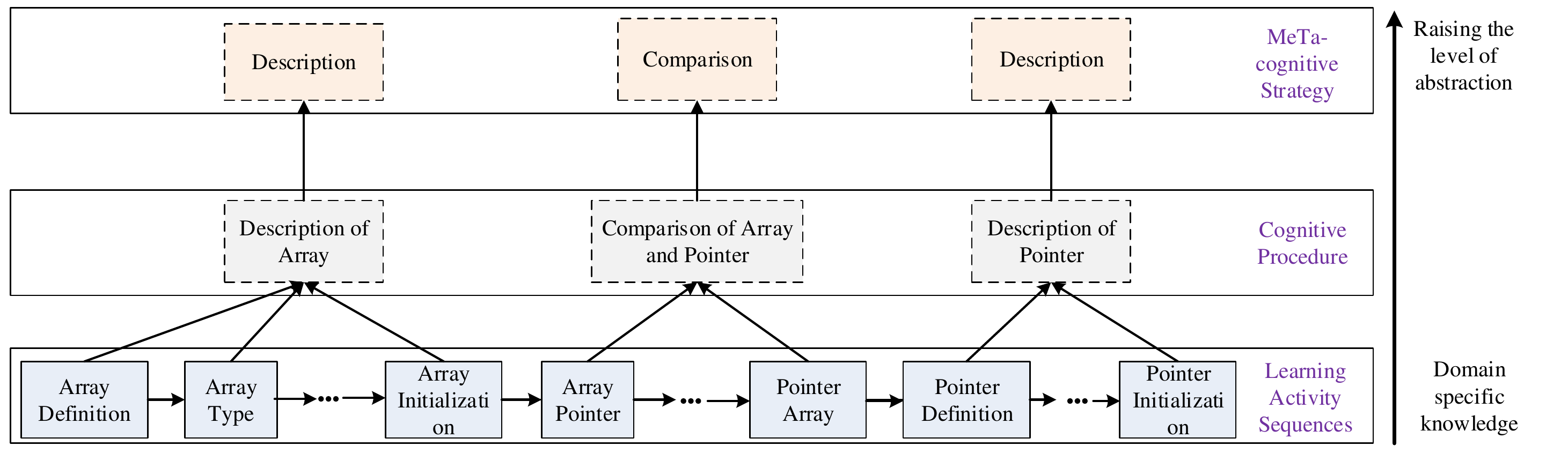}
	\caption{Detecting e-learners' cognitive and metacognitive strategy framework in actual question-solving scenario.}
	\label{fig_1}
\end{figure*}

\subsection{Modeling and Detecting Methods Review}

The traditional approaches to measuring learner's cognitive and metacognitive strategies are based on questionnaires. For example,  Zimmerman and Martinez-Pons \cite{7zimmerman1986development} conducted pilot interviews with high school students. They were asked to indicate in each topic which approaches they used to participate, to study, and to complete their assignments. Though the questionnaire approach that achieved 93\% accuracy using self-regulated learning (SRL) strategies, it failed to capture the dynamic and adaptive nature of SRL, as it did not consider the learning processes of students' knowledge-building, and problem-solving. Increasingly, researchers have focused more attention on analyzing the data collected from tracing students' learning process in a computer-based learning environment. For example, Perry and Winne \cite{19perry2006learning} categorized the activity traces into four statistical ways, which are frequency of study events, patterns of study activity, time and sequence of events, and content analysis of students' notes and summaries. Results suggest that analytic on trace data give better comprehension about how learners select, monitor, assemble, rehearse, and translate information. This provides useful raw materials for mapping SRL to learning effects.

In order to better understand learning process, data mining techniques have been adopted to measure learning or question-solving behaviors. Perera et al. \cite{27perera2009clustering} applied k-means clustering to find groups of similar teams and similar individual members, and then used sequential pattern mining algorithms to extract sequences of frequent events. Specially, Kinnebrew and Biswas have been researching metacognitive and SRL strategy in the past decade, which includes a series of methods to analyze learning behaviors. In 2011, they incorporated HMMs and exploratory data mining methodology for assessing and comparing students' learning behaviors from traces \cite{21kinnebrew2011comparative}. They employed a combination of sequence mining techniques to identify differential frequent patterns between groups of students in \cite{22kinnebrew2013contextualized}, analyzed the evolution of students' frequent behavior patterns over time in \cite{25kinnebrew2014analyzing}, used coherence analysis to characterize SRL behaviors in \cite{26segedy2015using}, developed a framework that combines model-driven strategy detection with data-driven pattern discovery for analyzing students' learning activity data in \cite{48J7368935}, etc. Segedy et al. \cite{23segedy2011modeling} proposed an integrated cognitive and metacognitive model for effective, self-regulated student learning in the Betty's Brain environment, while manually designing the mapping between the metacognitive and cognitive activities in their project that includes four aspects: goal-setting and planning, knowledge construction, monitoring and help-seeking. Particularly, in order to evaluate metacognition while individuals are learning a new task, Kim et al. \cite{kim2016applying} proposed a fuzzy linear regression model to analyze the relationship between retrospective confidence judgments (RCJ) and situation awareness (SA).

However, most of the existing researches seldom consider domain-specific knowledge constructions, they lack an automatic mapping mechanism between metacognitive strategies and activities,  or fail to make the cognitive and metacognitive process visualized.  These challenges have not been addressed. For example, the research from Kinnebrew et al. have a visual causal map that represents the relevant science phenomena as a set of entities connected by directed links that represent causal relations, called Betty's Brain \cite{2Leelawong2008Designing}. It, however, was unable to overcome the other two challenges \cite{49Riding2011The,50Rezaei2004Evaluation}.

In the constructivist's viewpoint, an integrated approach to the construction of procedural knowledge that should include discovery and acquisition of appropriate domain knowledge as well as subject choice, content scope selection, and sequence of learning. Therefore, the challenges of forming metacognition are not only to elicit procedural knowledge, but also to determine the domain scope and acquire appropriate knowledge.

With the development of knowledge engineering and educational psychology research, some recent outcomes, such as Thinking Maps (TM) and Knowledge Maps (KM), provide some basis of recognizing cognitive and metacognitive strategies. A knowledge map consists of the knowledge units of a given domain and the learning dependencies among them, which is a directed graph \cite{34Liu2012Topological} (a kind of domain knowledge structure). A KM is known to be correlated with a range of human endeavors \cite{35O2002Knowledge} and one emerging discipline that integrates different disciplines (e.g. economic, psychology, engineering etc.). Liu et al. had indicated that understanding the topological properties of knowledge maps may help gain better insights into human cognition structure and its mechanism \cite{34Liu2012Topological}. Only few research efforts utilizing KM to explore human cognition and metacognition process and the corresponding strategies, have been reported until now. Meanwhile, Thinking Maps claim to be a method to exploit higher order thinking skills and acts as a visual language to enhance learning experiences \cite{33Hyerle2014Thinking}. Therefore, the above research inspired us to combine KM with TM to explore the structure and process of e-learner's cognitive strategy by applying TM-based high-level abstraction in the context of KM. This could reveal how students use their cognitive and metacognitive strategies during question solving, especially in a kind of scenario of problem-based learning \cite{51Hmelo2004Problem}. This paper provides an effective framework for detecting cognitive and metacognitive strategy in the context of comparative question solving.

\section{Framework and Algorithm}
\label{Framework and Algorithm}
\subsection{Motivation}

Our proposed framework is shown in Fig.~\ref{fig_1}, which tries to explain how e-learners think, when solving questions, and illustrates our motivation inspired by a three-layer reverse engineering. The three-layer diagram consists of ``Learning Activity Sequences'' layer, ``Cognitive Procedure'' layer and ``Metacognitive Strategy'' layer. The hierarchical relationship from ``Learning Activity Sequences'' to ``Metacognitive Strategy'' abstractly describes a data-mining-based analysis process that maps specific behaviors and sequences of learning into cognitive elements and procedure by applying a bottom-up method. For  example, an e-learner's knowledge units of learning activity sequence is generated when solving the comparative question ``What are similarities and differences between array and pointer?'', \emph{Array Definition}, \emph{Array Type},..., \emph{2D Array Initialization}, etc., shown in Fig.~\ref{fig_1}, are recorded into a learning log, then a data-mining-based method on KMs and TMs analyzes these learning activity sequences and automatically maps them into different cognitive strategies, such as ``Description of Array'' and ``Comparison of Array and Pointer'', sequentially. This forms an instance of cognitive procedure for solving a comparative question, i.e. an instance of cognitive planning with specific knowledge units. After that, another method from the Thinking Map can be applied to map the cognitive procedure into the procedural cognitive knowledge comprised of ``Description-Comparison-Description'' without specific knowledge units, which is a metacognitive strategy the learner used before or stored in individual mind. Therefore, we can acquire the e-learner's cognitive and metacognitive strategy at different layers.

\begin{figure}[!t]
	\centering
	\includegraphics[width=1.8in]{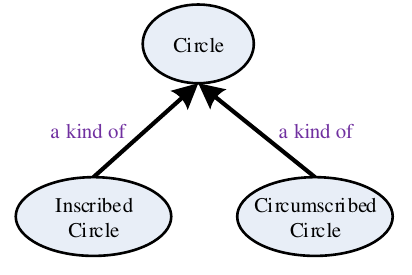}
	\caption{Semantic relation between knowledge units.}
	\label{fig_2}
\end{figure}

\subsection{Notation and Definition}

This section introduces notations and definitions used throughout this paper.

\newtheorem{mythm}{Definition}
\begin{mythm}[Knowledge Unit]\label{def_1:}
	A knowledge unit (denoted as $ku$) is the smallest integral learning object, such as a definition, a theorem or an algorithm \cite{36Zheng2011Yotta}. It is a node or entity in a graph, which usually composed of both name and content, and a core term of a $ku$ is its name. Note that one knowledge unit may only have one core term.
\end{mythm}

For example, a $ku$ \emph{Array Definition} in ``The C Programming Language'' course knowledge map can be described as ``Name: Array Definition; Content: An array is a group of variables of a type occupying a contiguous region of memory \cite{52bailey2005introduction}'', and the core term of the knowledge unit is ``array definition''.

\begin{mythm}[Semantic Relation Between Two Kus]\label{def_2:}
	A semantic relation (denoted as $r$) is a directed edge in graph/map, which indicates a relationship from one $ku$ to another $ku$.
\end{mythm}

The type of semantic relation is diverse, commonly occurring are: ``a part of'', ``an attribute'', ``a definition'', ``a kind of'', ``a type of'', ``an association'', ``similar to'', ``an initial cause'', and ``a result'', etc. For example, as shown in Fig.~\ref{fig_2}, a circle can be divided into two different types: inscribed circles and circumscribed circles. Therefore, the semantic relation between \emph{Circle} $ku$ and \emph{Inscribed Circle} $ku$ (or \emph{Circumscribed Circle} $ku$) can be called ``a kind of''.

\begin{figure}[!t]
	\centering
	\includegraphics[width=2.8in]{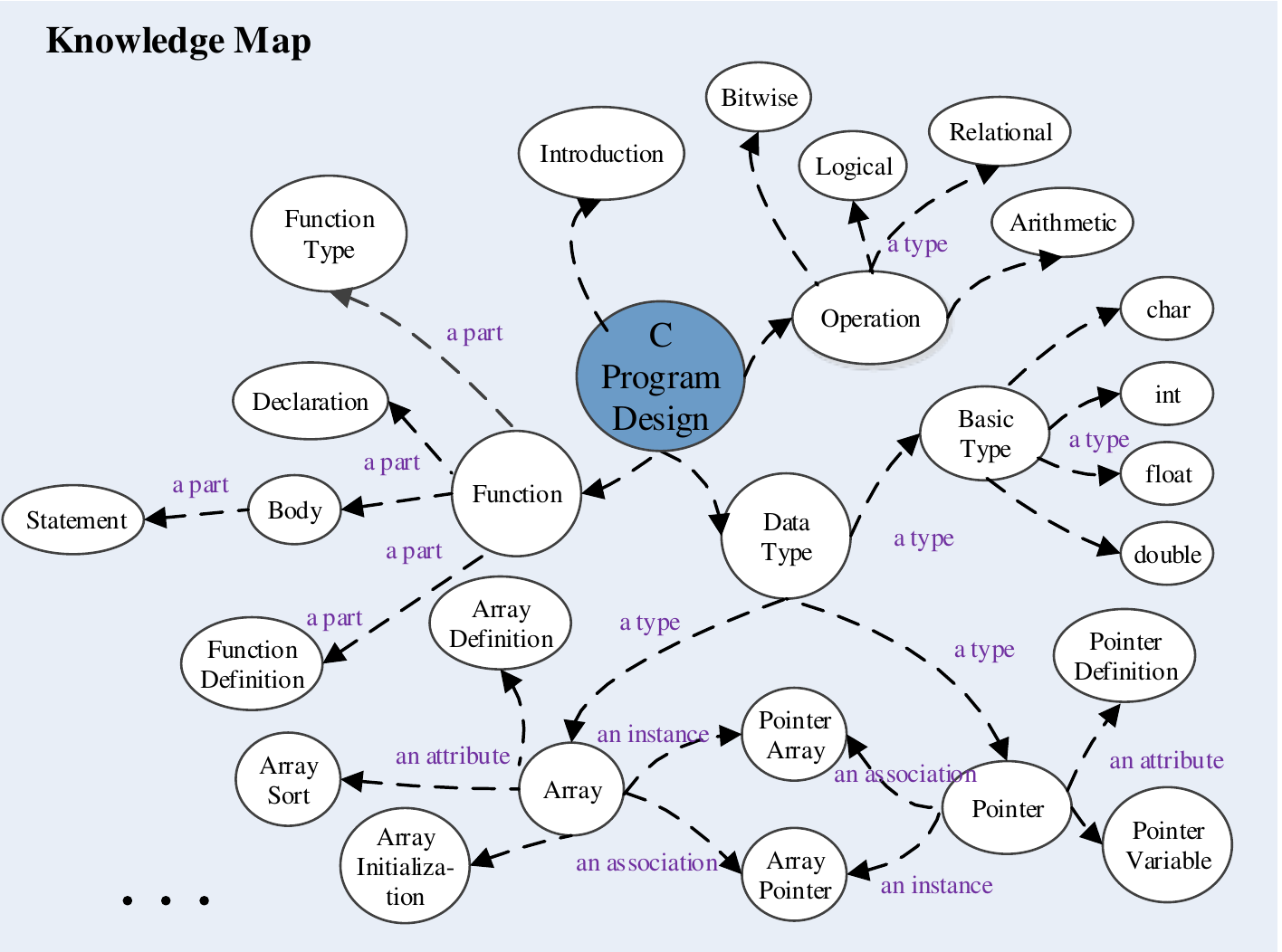}
	\caption{A partial knowledge map of ``The C Programming Language'' course.}
	\label{fig_3}
\end{figure}

\begin{mythm}[Knowledge Map]\label{def_3:}
	A knowledge map is a knowledge structure of a course or discipline organized and managed its knowledge unit in a graph-based manner \cite{34Liu2012Topological} (denoted as KM). A KM is a directed graph composed of entities (nodes) and relations (edges). Besides, it can be viewed as a semantic network, in which each node is a knowledge unit and edge is represented as a triple of the form \emph{(head $ku$, semantic relation, tail $ku$)}, denoted as $ke$. Formally, we define a KM: $KM=(KU,KE)$, in which, $ku\in KU$ and $ke\in KE$.
\end{mythm}

Fig.~\ref{fig_3} depicts a partial KM of ``The C Programming Language'' course, in which, each edge has a semantic relation, for example, the edge between \emph{Function} and \emph{Body} will be represented as \emph{(Function, ``a part'', Body)}.

\begin{mythm}[Learning Activity Sequence]\label{def_4:}
	A learning activity sequence obtained from a learning log reflects a track of a student's learning process during a period, which is an output of the first layer in Fig.~\ref{fig_1}. A $LAS$ containing $n$ $kus$ can be represented as follow:
	\begin{equation*}
	LAS=<ku_1,ku_2,\ldots,ku_n>
	\end{equation*}
\end{mythm}

\begin{figure}[!t]
	\centering
	\includegraphics[width=2.5in]{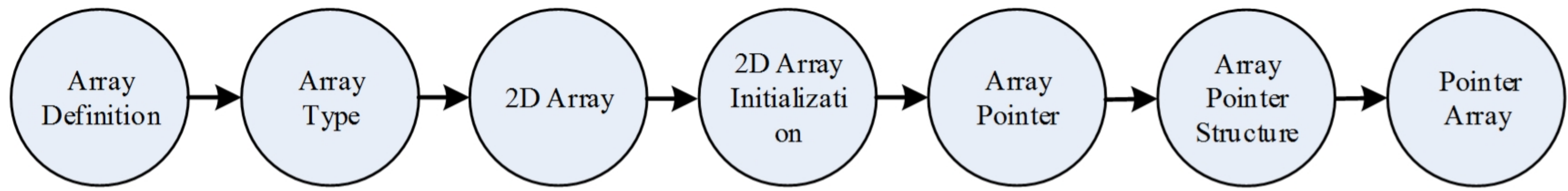}
	\caption{An instance of learner's learning activity sequence.}
	\label{fig_4}
\end{figure}

For example, an instance of a specific student's $LAS$ may be \emph{$<$Array Definition, Array Type, 2D Array, 2D Array Initialization, Array Pointer, Array Pointer Structure, Pointer Array$>$} to answer ``What are similarities and differences between array and pointer?'' question, shown in Fig.~\ref{fig_4}.

\begin{mythm}[Thinking Map]\label{def_5:}
	Thinking Map (denoted as TM) is a common visual language for learning communities to describe and visualize the thinking process \cite{33Hyerle2014Thinking}.
\end{mythm}

\begin{figure}[!t]
	\centering
	\includegraphics[width=3.5in]{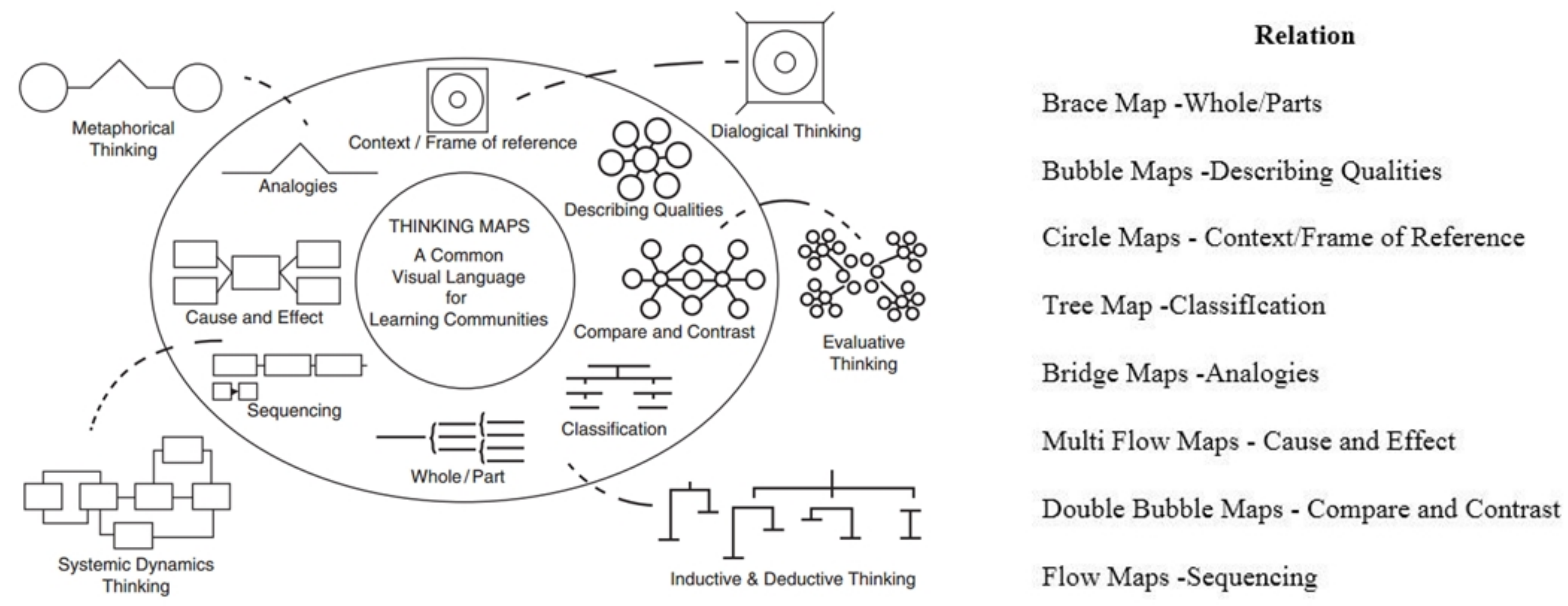}
	\caption{Thinking Map \cite{33Hyerle2014Thinking}.}
	\label{fig_5}
\end{figure}

Fig.~\ref{fig_5} shows eight kinds of graph methods: \emph{Brace Map, Bubble Map, Circle Map, Tree Map, Bridge Map, Multi-Flow Map, Double Bubble Map and Flow Map}. Each map represents a basic thinking process when thinking about questions, and the corresponding eight thinking processes are \emph{Whole/Part, Describing Qualities, Context/Frame of Reference, Classification, Analogies, Cause and Effect, Compare and Contrast and Sequencing}.

A KM describes the knowledge units and their semantic relationships within a specific domain. While each thinking map means nothing if the corresponding process does not include the specific knowledge, especially when a person is thinking about how to answer the domain specific questions. Hence, the methodology of combining KMs with TMs can model a person's thinking process in solving questions, when they ponder about them.

To deal with this, there are three things that need to be considered: questions, mapping from questions to KM, and mapping from question-oriented KM to TM.

\begin{figure}[!t]
	\centering
	\includegraphics[width=3.5in]{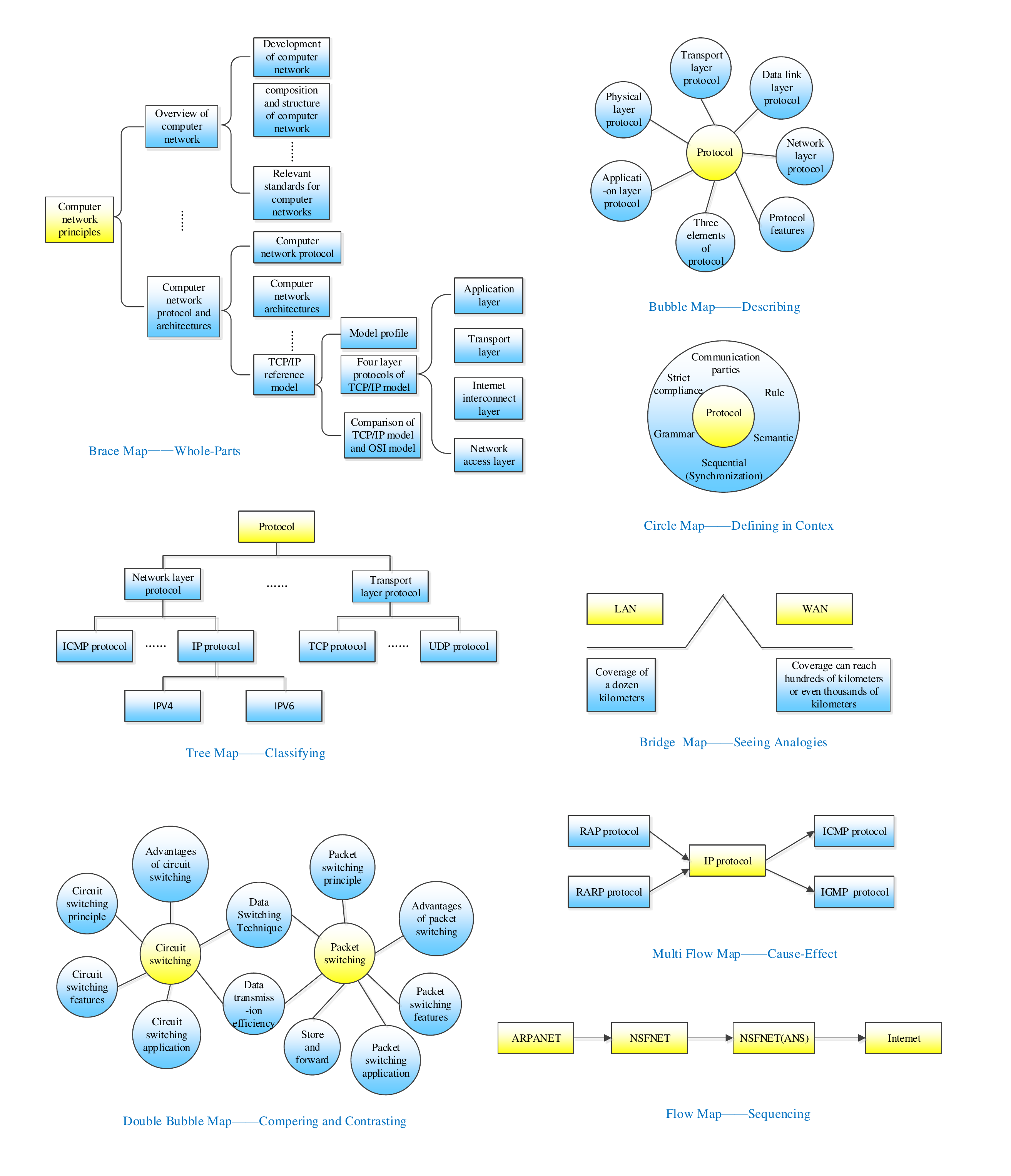}
	\caption{Thinking Maps with specific questions in the context of ``Computer Network Principle'' KM.}
	\label{fig_6}
\end{figure}

In this paper, we concentrate on how e-learners solve problems by using their cognitive and metacognitive strategies. To gain better understanding this research and its result, we adopt less complex questions to test e-learners. For mapping from questions to KM, the core terms will help to locate the corresponding knowledge units, according to Definition \ref{def_6:}. For mapping from question-oriented KM to TM, we utilize a mapping relationship between KM and TM. Particularly, in Fig.~\ref{fig_6}, we illustrated the eight kinds of Thinking Maps with specific questions in the context of the KM of ``Computer Network Principle'' course.

Take the Bubble Map in Fig.~\ref{fig_6} as an example. Given a question, such as ``How do you know about protocol?'', then a learner may invoke the knowledge related to ``protocol'' and construct the bubble-like map. In this scenario, firstly, abstract the core term ``protocol'' from the question; secondly, extract the knowledge units corresponding to the ``protocol'' course located in KM; thirdly, mine or find a specific knowledge submap in the KM by using the relation mapping between KM and TM. In this process, we call knowledge unit \emph{Protocol} a core knowledge unit, and the specific knowledge submap mined in KM is deem to a thinking map with core knowledge units.

Based on the above, we coin some important definitions as followings.

\begin{mythm}[Core knowledge Unit]\label{def_6:}
	A core knowledge unit (denoted as $cku$) is a centric phrase of a question sentence, which builds a bridge between a question and knowledge units.
\end{mythm}

For example, given a specific question, ``What is the definition of array?'', it can be inferred that ``array'' is the core item and ``definition'' denotes an attribute of ``array'', so that, we can call \emph{Array} is a core knowledge unit, and \emph{Array Definition} is a subordinate $ku$ of \emph{Array}. Similarly, the core knowledge units of this question ``What are similarities and differences between array and pointer?'' are \emph{Array} and \emph{Pointer}. The $ckus$ are all colored in yellow this paper.

\begin{mythm}[Thinking Maps with $cku$]\label{def_7:}
	A Thinking Map with $cku$ is a special knowledge submap, which can be found in a domain-specific KM by the presented algorithm, in Section III-C.
\end{mythm}

For instance, we can follow the ``a type of'' or ``a kind of'' semantic relation to construct a Tree Map with \emph{Protocol} $cku$ submap, and follow the ``an association'' semantic relation to build a Circle Map with \emph{Protocol} $cku$ submap, depicted in Fig.~\ref{fig_6}. Consequently, more than one submap is built with the \emph{Protocol} $cku$. Moreover, We provide a description of TM with $cku$ submap searched algorithm discussed in Algorithm \ref{alg_2}.

Particularly, we introduce two specific knowledge submaps: Descriptive Knowledge Submap and Connective Knowledge Submap, which are the basic elements of Bubble Map and Double Bubble Map with $cku$, the detailed description of these submaps are as follow.

\begin{figure}[!t]
	\centering
	\subfloat[Descriptive knowledge submap of \emph{Array}]{\includegraphics[width=1.1in]{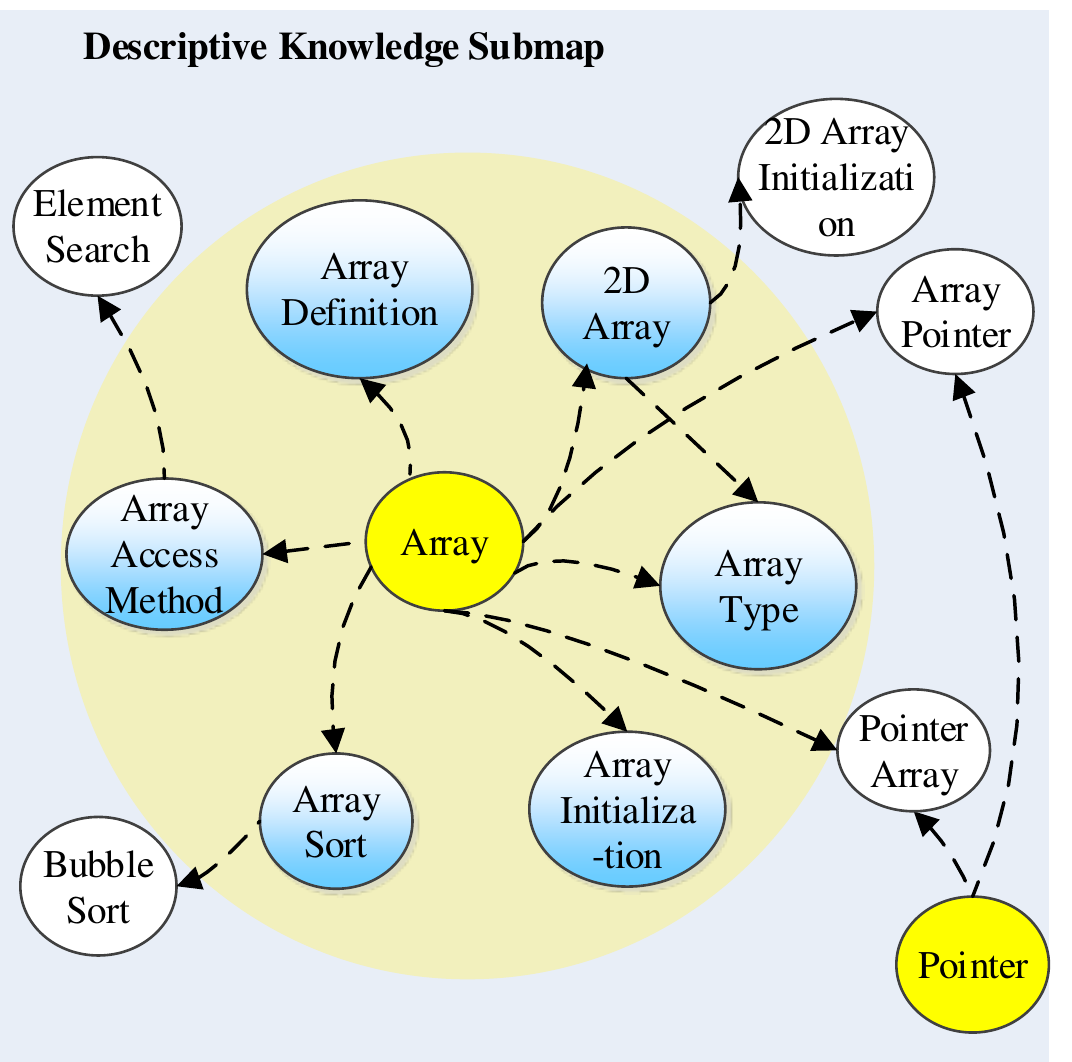} \label{fig_7a}}
	\subfloat[Connective Knowledge Submap of \emph{Array} and \emph{Pointer}]{\includegraphics[width=1.1in]{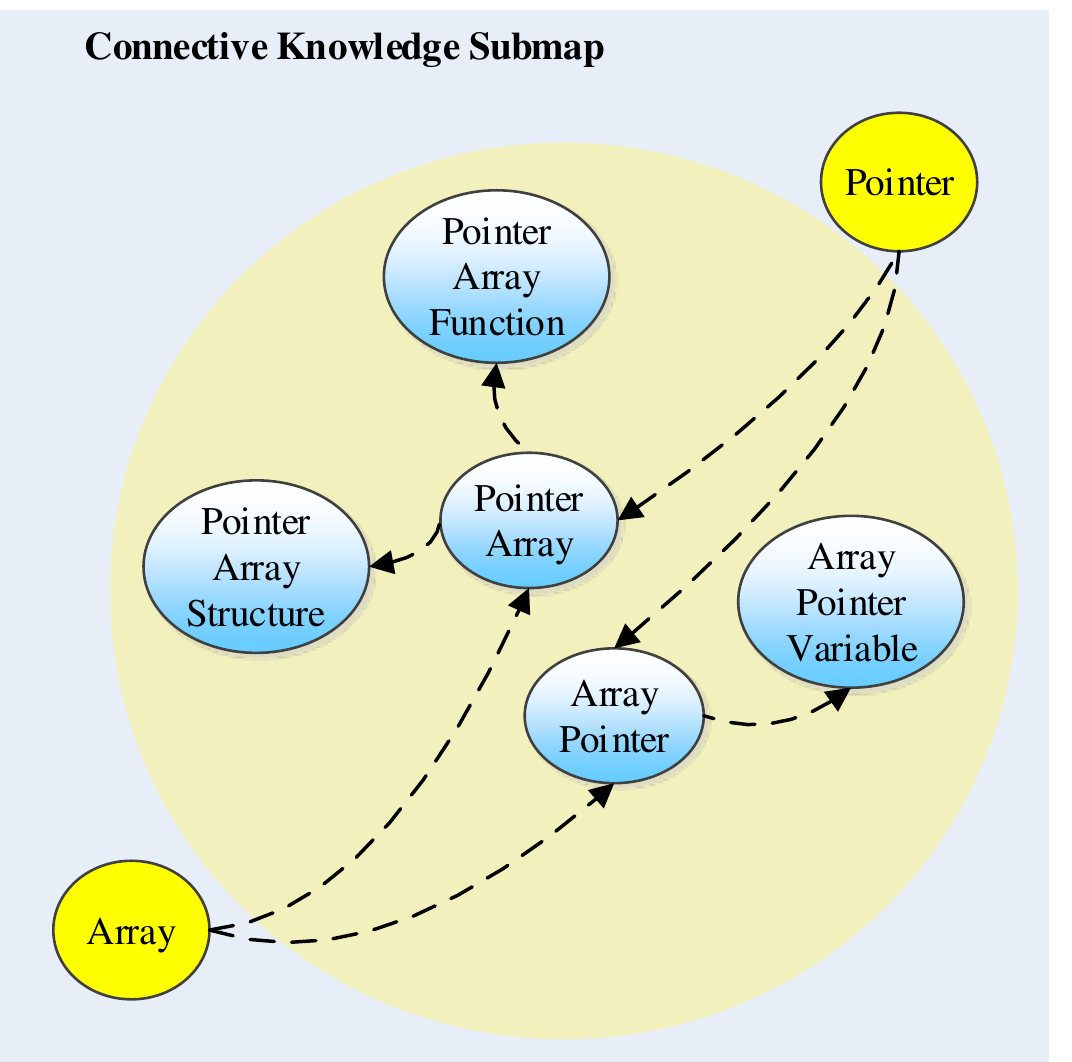} \label{fig_7b}}
	\subfloat[Descriptive knowledge submap of \emph{Pointer}]{\includegraphics[width=1.1in]{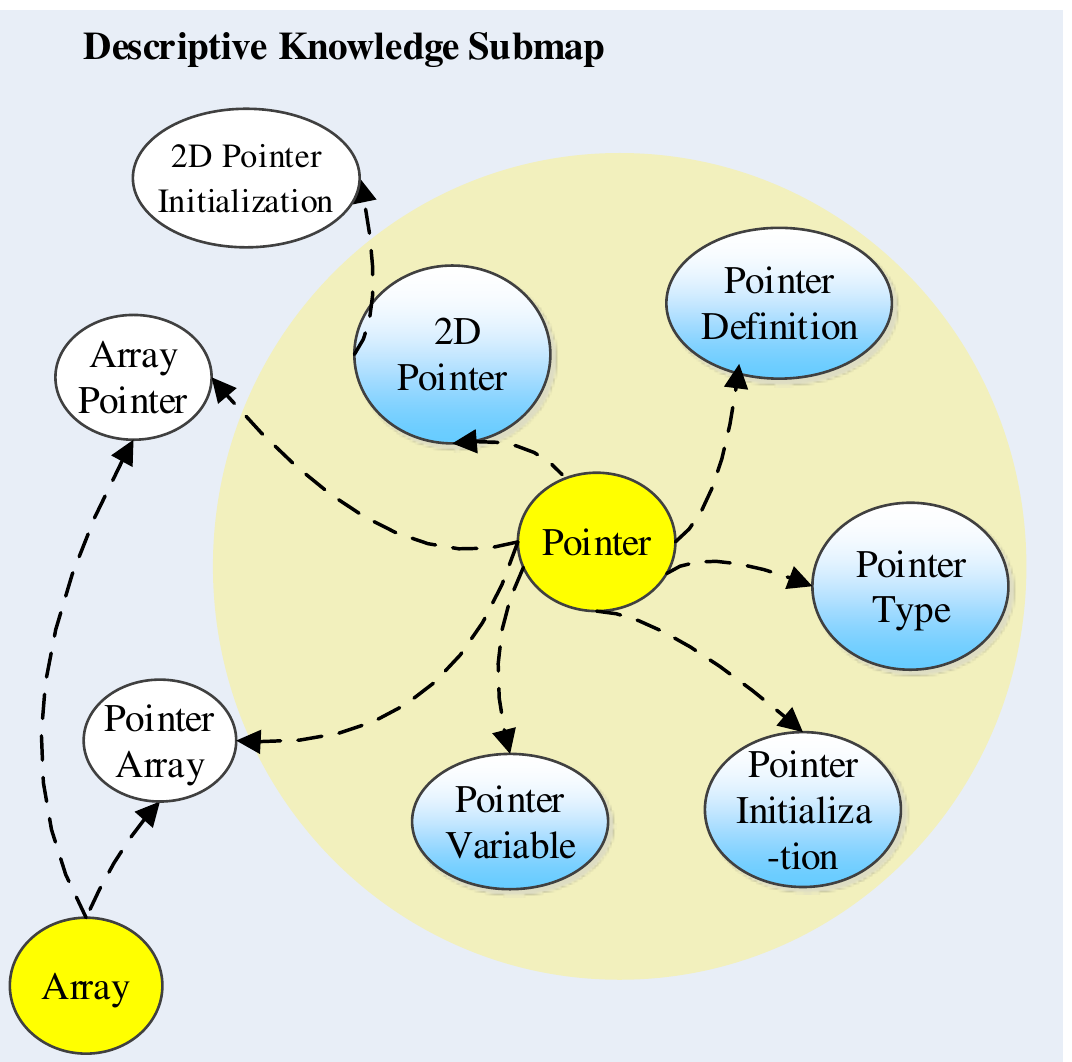} \label{fig_7c}}\\
	\subfloat[Double Buble Map with \emph{Array} and \emph{Pointer}]{\includegraphics[width=3in]{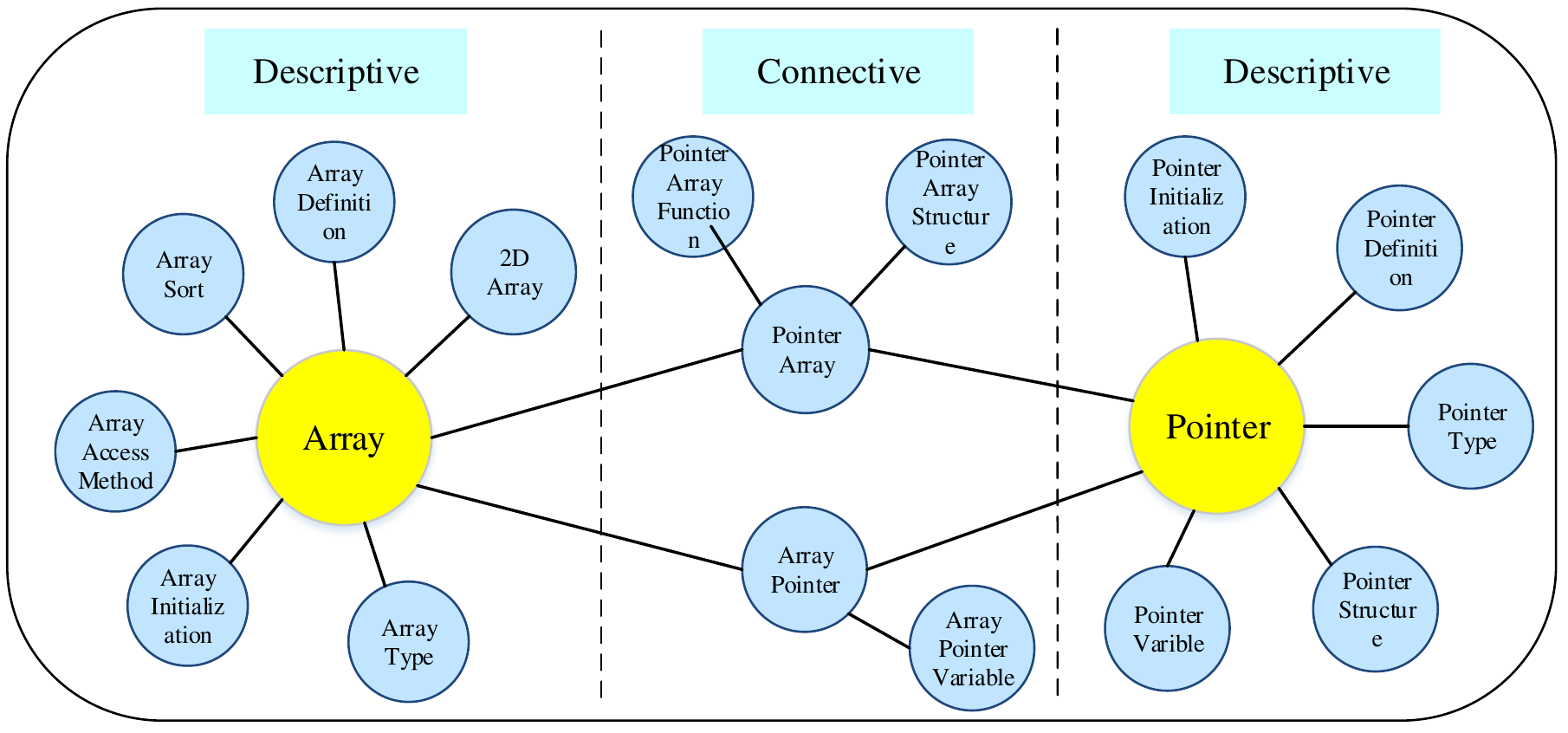} \label{fig_7d}}%	
	
	\caption{Knowledge submaps searched in KM.}
	\label{fig_7}
\end{figure}

A Descriptive Knowledge Submap can be defined by:
\begin{equation*}
KM^{descriptive}_{sub}=(KU^{desc}_{sub},KE^{desc}_{sub})
\end{equation*}
Where $KU^{desc}_{sub}$  is a set of $kus$, which are first-order neighbor vertices of a $cku$ colored in blue, and $KE^{desc}_{sub}$ is a set of semantic relations between $cku$ colored in yellow and other $ku$. That is, it satisfies $\forall ku\in KU^{desc}_{sub},(cku,r,ku)\in KE^{desc}_{sub}$. For example, we searched two submaps from ``The C Programming Language'' course KM, which are descriptive knowledge submap of \emph{Array} $KM^{desc(array)}_{sub}$ and descriptive knowledge submap of \emph{Pointer} $KM^{desc(pointer)}_{sub}$, illustrated within the light yellow area of Fig.~\ref{fig_7a} and \ref{fig_7c}, and the overrange $kus$ are colored in white.

Similarly, a Connective Knowledge Submap can be defined by:
\begin{equation*}
KM^{connective}_{sub}=(KU^{conn}_{sub},KE^{conn}_{sub})
\end{equation*}
A connective knowledge submap usually contains two core knowledge units mapped by core terms, such as \emph{Array} and \emph{Pointer} units colored in yellow, as shown in the connective knowledge submap of \emph{Array} and \emph{Pointer} $KM_{sub}^{conn(array,pointer)}$, Fig.~\ref{fig_7b}. In which, the connective $kus$ of two $ckus$ are \emph{Array Pointer} and \emph{Pointer Array}.

To integrate descriptive submap and connective submap, i.e., merging the \emph{Array} descriptive knowledge submap, the \emph{Pointer} descriptive knowledge submap and the connective knowledge submap of the two $kus$, we can obtain a Double Bubble Map with \emph{Array} and \emph{Pointer}, exhibited in Fig.~\ref{fig_7d}, which illustrates the similarities and differences between array and pointer.

Next, we will proposed Cognition Control Measure, Cognition Control Sequence, Cognitive Procedure and Metacognitive Strategy Sequence, to analysis the learning activity sequence and measure the individual question-solving process.

\begin{mythm}[Cognition Control Measure]\label{def_8:}
	Cognition control measure (denoted as $CCM$) is a coverage rate of knowledge submaps, which provides a quantitative calculation of cognition. The definition is as below:
	\begin{equation*}
	CCM=(\frac{N_{1}}{|KU_{sub1}|-k},\frac{N_{2}}{|KU_{sub2}|-k},\dots,\frac{N_{i}}{|KU_{subi}|-k})
	\end{equation*}
\end{mythm}

Where, $N_{i}$ denotes the number of $ku$ which belong to the responding $KM_{subi}$  at a learning moment; $|KU_{subi}|$ denotes the numbers of units in $KM_{sub}$; $k\in \{1,2\}$, denotes the number of core knowledge units in its submap; $i$ denotes the number of searched knowledge submaps, in addition, the type and order of these submaps can be extracted from learning activity sequence.

Take the $LAS$ in Fig.~\ref{fig_4} as an example, assuming that the corresponding knowledge submaps are searched and represented in Fig.~\ref{fig_7a}, \ref{fig_7b} and \ref{fig_7c}. As can be seen, the number of $kus$ in $KM^{desc(array)}_{sub}$ is 6 (not including \emph{Array} $cku$), and there are 3 $kus$ in current learning activity sequence, so the knowledge unit coverage rate of $KM^{desc(array)}_{sub}$ is: $RatioSub_{desc(array)}=\frac{1}{2}$ . Similarly, the number of $kus$ in $KM^{conn(array,pointer)}_{sub}$ is 5 (not including \emph{Array} $cku$ and \emph{Pointer} $cku$), and there are 3 $kus$ in current learning activity sequence, so the knowledge unit coverage rate of $KM^{conn(array,pointer)}_{sub}$ is: $RatioSub_{conn(array,pointer)}=\frac{3}{5}$. However, there are no learning activities in the Knowledge Submap describing \emph{Pointer}, so the knowledge unit coverage rate of \emph{Pointer} $KM^{desc(pointer)}_{sub}$  is: $RatioSub_{desc(array)}=0$ . Hence, the $CCM$ of this learning sequences in Fig.~\ref{fig_4} is $(\frac{1}{2},\frac{3}{5},0)$.

\begin{mythm}[Cognition Control Sequence]\label{def_9:}
	Cognition Control Sequence (denoted as $S_{cog}$ ) reflects the dynamic process of completing a certain task, hence, a $S_{cog}$ consists of $CCMs$ of different moments. The representation of $S_{cog}$  is given below:
	\begin{equation*}
	S_{cog}=<CCM_{1},CCM_{2},\dots,CCM_{n}>
	\end{equation*}
\end{mythm}

\begin{mythm}[Cognitive Procedure]\label{def_10}
	Cognitive Procedure also called cognitive planning with specific knowledge units. It reflects the dynamic changes of the learners' cognitive strategies, which is an output of the second layer in Fig.~\ref{fig_1}.
\end{mythm}

Based on TM with $cku$ and the $S_{cog}$, we can abstracted that ``Description of Array'' and ``Description of Pointer'' are specific cognitive strategies depicted by Bubble Maps with \emph{Array} and \emph{Pointer} respectively, and ``Comparison of Array and Pointer'' cognitive strategy is depicted by a Double Bubble Map with the two core knowledge units. 

\begin{mythm}[Metacognitive Strategy Sequence]\label{def_11}
	A metacognitive strategy sequence is represented by:
	\begin{equation*}
	S_{meta-cog}=(cog_{1},cog_{2},\dots,cog_{n})
	\end{equation*}
\end{mythm}

Where, $cog$ is the selected cognitive strategy, and the order of sequence is combinations of cognitive strategies. $S_{meta-cog}$  is the output of the third layer in our framework. From which, generally under the scenario of solving a complex question, a combination of different metacognitive strategies can be mined through further abstraction. Therefore, the metacognitive strategy of solving ``What are similarities and differences between array and pointer?'' may be represented as \emph{(Description, Comparison, Description)}.

\subsection{The Proposed Pattern Mining Algorithm}

This section presents our cognitive strategy and metacognition strategy patterns mining algorithm under the scenario of question solving. Algorithm \ref{alg:mining} describes a main structure of our mining algorithm, which follows the reverse engineering engine idea to raise the abstraction level from learning activities to metacognitive strategy pattern. Algorithm \ref{alg_2} finds the candidate submaps through a heuristic search method, which is guided by Thinking Map. Given specific core knowledge units, the corresponding candidate submaps with $cku$ will be found. Algorithm \ref{alg_3} transforms a three-dimensional vector into a one-dimensional sequence, to reduce the time and space complexity. Algorithm \ref{alg_4} encodes the knowledge unit coverage rate and Algorithm \ref{alg_5} decodes the Cognition Control Sequence.

\begin{algorithm}  
	\caption{Cognitive Strategy and Metacognitive Strategy Pattern Mining.}  
	\label{alg:mining}  
	\begin{algorithmic}[1]
		\REQUIRE  Paramset \textbf{\{courseId, CI, k-depth\}}, in which,  \textbf{courseId} points to a specific knowledge map which the question belongs to; \textbf{CI} refers to a core item set; \textbf{k-depth} denotes the max hop visited in KM. Note that \textbf{courseId} and \textbf{CI} are extracted from the given question for solving.
		\ENSURE  Cognitive Strategy\\
		Metacognition Strategy Pattern: $\{S_{meta-cog}\}$;
		\STATE Acquire a related domain-specific  according to $courseId$.
		\STATE Mapping $CI$ into $CKU$  /*$CKU=\{cku\}$ or $CKU=\{cku_{1}, cku_{2}\}$*/
		\WHILE {$(i>n)$}
		\STATE find $cku_{i}$ in $KM$
		\ENDWHILE
		\STATE Call $SubmapSearch(CKU, k-depth)$ to obtain the candidate submaps
		\FOR {each $LAS_{i}$ in $\{LAS\}$}
		\STATE Remove the irrelevant submaps
		\STATE Calculate $CCM_{i}$
		\ENDFOR
		\STATE Generate the cognition control sequence: $S_{cog}=<CCM_{1},CCM_{2},\dots,CCM_{n}>$
		\STATE Obtain the Cognitive Strategy Sequence with cku:$(Cognitive-Strategy(cku)_{1}, Cognitive-Strategy(cku)_{2},\dots, Cognitive-Strategy(cku)_{i} ,\dots)$
		\FOR {$S_{cog}$ in $\{S_{cog}\}$}
		\STATE $\hat{S_{cog}}=EncodeCogSeq(S_{cog})$
		\ENDFOR
		\STATE $\{\hat{Pattern_{cog}}\}=GSP(\hat{S_{cog}},minsup)$
		\STATE $\{Pattern_{cog}\}=DecodeCogPattern(\hat{Pattern_{cog}})$
		\STATE Obtain $\{S_{meta-cog}\}$ from $\{Pattern_{cog}\}$ through raising the level of abstraction
		\RETURN $\{S_{meta-cog}\}$
	\end{algorithmic}
\end{algorithm}

In Algorithm \ref{alg:mining}, Step 1 finds the domain Knowledge Map based $courseId$ that points to a specific knowledge map; Step 2 to 5 aim to search the core knowledge unit in KM mapped by $CI$ representing a core item set; Step 6 searches the candidate submaps by calling  $SubmapSearch$ algorithm presented in Algorithm \ref{alg_2}; Step 7 to 11 generates the cognition control sequence by calculating knowledge unit coverage rate in $KM_{sub}$, and get rid of the irrelevant submaps, so that we obtain the cognitive strategy sequence in step 12; Step 13 to 15 encode the dimensional cognition control sequence described specifically in  Algorithm \ref{alg_3} $EncodeCogSeq$; Step 16 mines frequent patterns from the encoded cognition control sequence by the sequential pattern mining algorithm \emph{GSP} (Generalized Sequential Patterns) \cite{53Srikant1996Mining}. From the definition of knowledge unit coverage rate, the cognition control sequence of comparison question is a three-dimensional vector, we transform the cognition control sequence into a one-dimensional sequence defined by the algorithm in Algorithm \ref{alg_4} $Seg$; Step 17 decodes the frequent patterns to get the metacognition strategy pattern in Algorithm \ref{alg_5} $DecodeCogPattern$. Therefore, we obtain the metacognition strategy pattern in the last Step.

\begin{algorithm}[h]  
	\caption{Heuristic Submaps Search in Thinking Maps Algorithm $SubmapSearch$.} 
	\label{alg_2}  
	\begin{algorithmic}[1]
		\REQUIRE  Paramset \textbf{\{CKU, k-depth\}}  
		\ENSURE  Candidate knowledge submaps set: $\{KM_{sub}\}$
		\STATE $KU_{sub}\xleftarrow{} \{\}$, $KE_{sub}\xleftarrow{} \{\}$
		\IF{$|CKU|==1$}
		\STATE $cku\xleftarrow{} CKU[0]$
		\WHILE {hop$\leq$1}
		\STATE Find candidate $ku_{i}$ through semantic relation $r$ start from $cku$
		\STATE Save $ku_{i}$ into $KU_{sub}$, and save $(ku_{i},r,ku_{j})$ into $KE_{sub}$
		\ENDWHILE
		\STATE Obtain $KU_{sub}=\{ku_{1},ku_{2},\dots, ku_{n}\}$, $KE_{sub}=\{(cku,r,ku_{m},(ku_{n},r,cku),\dots)\}$, $m,n=1,2,\dots$
		\STATE Repeat step 5-6 to construct the one hop submap of $cku$, $KM_{Bub}\xleftarrow\ r_{an\ attribute}$, $KM_{Cir}\xleftarrow\ r_{an\ association}$, $KM_{Mf}\xleftarrow\ r_{an\ cause/result}$, thus obtain Bubble Map, Circle Map and Multi-Flow\ Map with $cku$ submaps.
		\STATE Repeat step 5-6 to construct the $k-depth$ hop submap of $cku$, $KM_{Tree}\xleftarrow\ r_{an\ type}$, $KM_{Brace}\xleftarrow\ r_{an\ part}$ thus obtain Tree Map and Brace Map with $cku$ submaps.
		\RETURN $\{KM_{Bub}, KM_{Cir}, KM_{Mf}, KM_{Tree}, KM_{Brace}\}$
		\ELSE
		\STATE $cku_{1}\xleftarrow{} CKU[0]$, $cku_{2}\xleftarrow{} CKU[1]$
		\STATE Repeat step 5-6 to obtain $KM_{Bub}(cku_{1})$ and $KM_{Bub}(cku_{2})$
		\STATE Obtain part of connective submap with two $ckus$, $\hat{KM_{conn}}=KM_{Bub}(cku_{1})\cap KM_{Bub}(cku_{2})$
		\STATE Build $\{kus_{conn}\} = KU_{Bub}(cku_{1})\cup KU_{Bub}(cku_{2})$
		\FOR {each $ku_{i}$ in $\{kus_{conn}\}$}
		\STATE Repeat step 5-6 to Obtain $KM_{i}(ku_{i})$ 
		\STATE $\hat{KM_{conn}}=\hat{KM_{conn}}\cup KM_{i}(ku_{i})$
		\ENDFOR 
		\STATE Obtain connective submap with two $ckus$, $KM_{conn}(cku_{1},cku_{2})=\hat{KM_{conn}}$
		%		\STATE Obtain Double Bubble Map with $cku$ submap $KM_{D-bub}=KM_{Bub}(cku_{1})\cup KM_{conn}(cku_{1},cku_{2})\cup KM_{Bub}(cku_{2})$
		\RETURN $KM_{conn}(cku_{1},cku_{2})$
		\ENDIF		
	\end{algorithmic}
\end{algorithm}

\begin{figure}[!t]
	\centering
	\subfloat[“The C Programming Language” course.]{\includegraphics[width=1.7in]{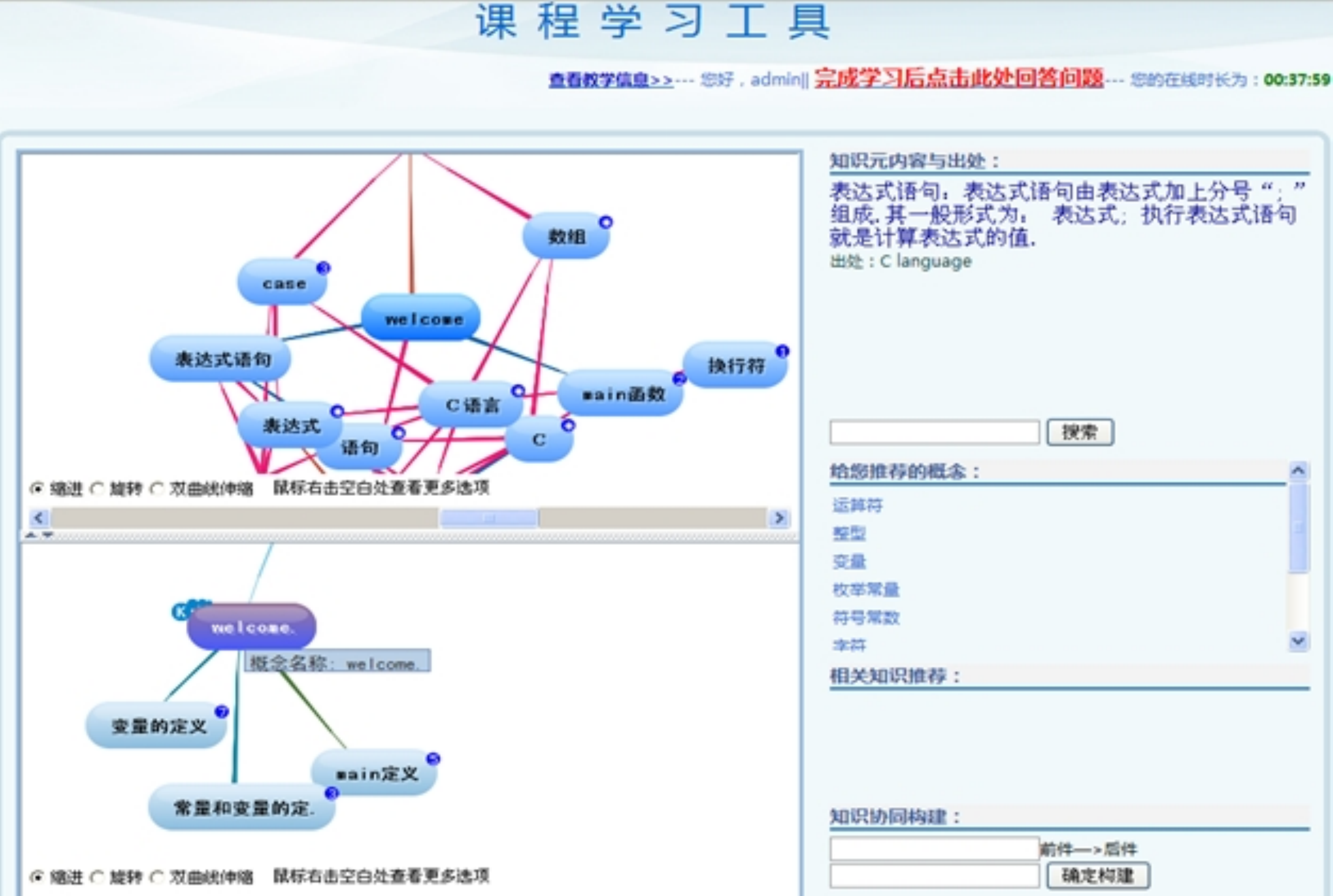} \label{fig_8a}}
	\subfloat[“Computer Network Principle” course.]{\includegraphics[width=1.7in]{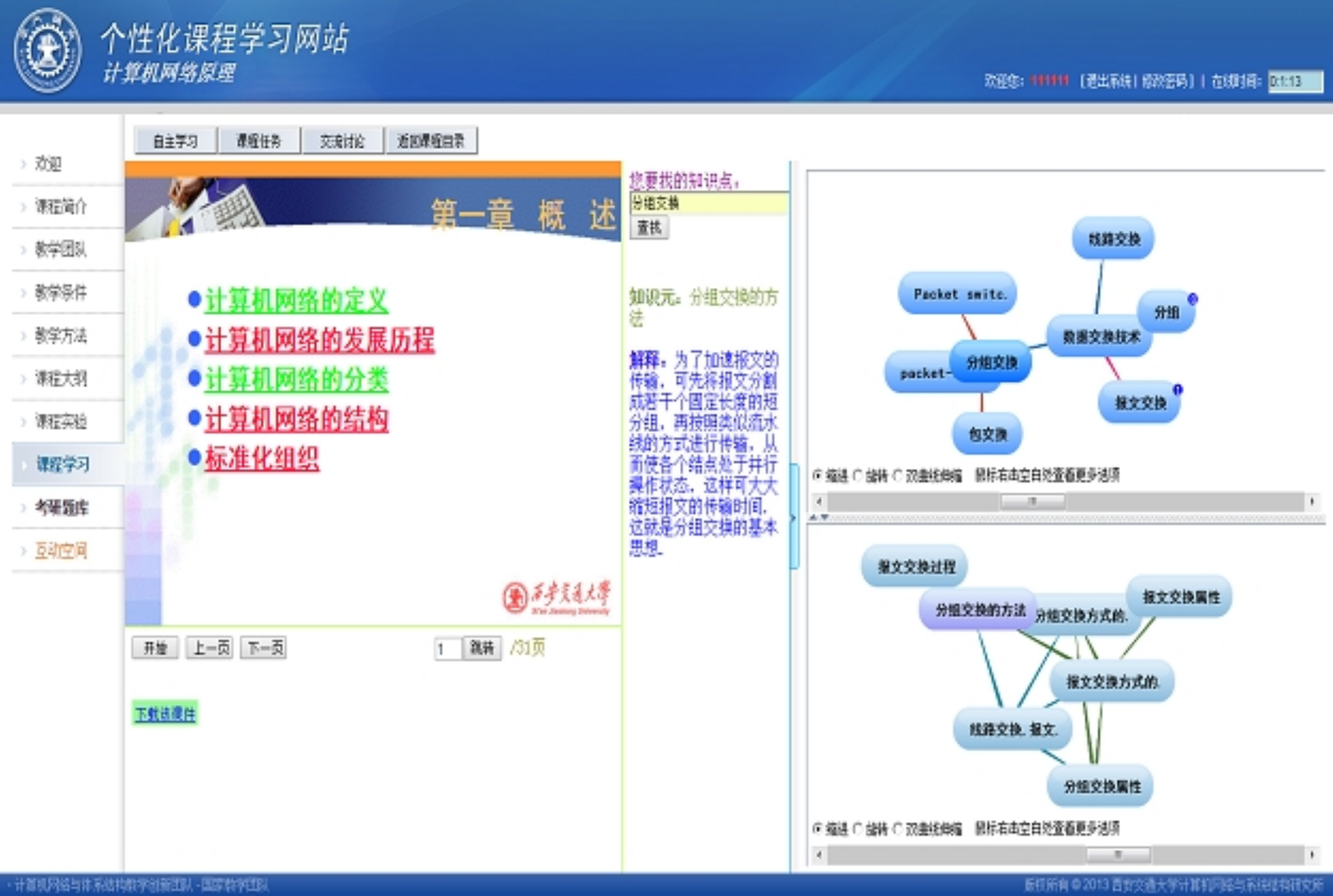} \label{fig_8b}}
	\caption{Online learning prototype system.}
	\label{fig_8}
\end{figure}

Algorithm 2 detailedly describes the process of finding relevant submaps with specific $cku$ from the eight Thinking Maps defined in Definition \ref{def_5:}.

\begin{algorithm}[h]
	\caption{Cognition Control Sequence Encoding Algorithm $EncodeCogSeq$.}  
	\label{alg_3}  
	\begin{algorithmic}[1]
		\REQUIRE  Dimensional cognition control sequence:\\ $S_{cog}=<CCM_{1},CCM_{2},\dots,CCM_{n}>$
		\ENSURE  One-dimensional sequence:\\ $\hat{S_{cog}}=<\hat{CCM_{1}},\hat{CCM_{2}},\dots,\hat{CCM_{n}}>$
		\FORALL {$CCM_{i}$ in \{$S_{cog}$\}}
		\FOR {$x_{i,j}$ in $CCM_{i}$}
		\STATE \[ f(x_{i,j})=
		\begin{cases}
		1 & x_{i,j}=1\\
		0 & x_{i,j}=0\\
		0.5 & x_{i,j}\in (0,1)
		\end{cases}\]
		\STATE $\hat{x_{i,j}}=f(x_{i,j})$
		\STATE $\hat{CCM}=\sum_{j=1}^{n}\hat{x_{i,j}}*10^{7-2j}$
		\ENDFOR
		\STATE $\hat{S_{cog}}.add(\hat{CCM_{i}})$
		\ENDFOR
		\RETURN $\hat{S_{cog}}$
	\end{algorithmic}
\end{algorithm}

\begin{algorithm}[h]
	\caption{Cognition Control Sequence Decoding Algorithm $DecodeCogPattern$.}  
	\label{alg_4}  
	\begin{algorithmic}[1]
		\REQUIRE  One-dimensional sequence: $\hat{S_{cog}}$
		\ENSURE  Strategy pattern sequence: $\{Pattern_{cog}\}$
		\FOR {$ \hat{CCM_{i}}\ in \ \hat{S_{cog}} $}
		\STATE $\hat{X_{i}}=Seg(\hat{CCM_{i}})$
		\STATE $CCM_{i}=\hat{X_{i}}/100$
		\ENDFOR
		\STATE $S_{cog}.add(CCM_{i})$
		\RETURN $S_{cog}$
	\end{algorithmic}
\end{algorithm}

\begin{algorithm}[h]
	\caption{Encoded Knowledge Unit Coverage Rate Segment Algorithm $Seg$.}  
	\label{alg_5}  
	\begin{algorithmic}[1]
		\REQUIRE Encoded knowledge unit coverage rate: $\hat{CCM_{i}}$  
		\ENSURE  Truncated vector: $\hat{X_{i}}=(\hat{x_{i,1}},\hat{x_{i,2}},\dots,\hat{x_{i,n}})$
		\STATE $N=|max\ \hat{CCM_{i}}|$
		\STATE $j\xleftarrow{} 1$
		\WHILE {$j\leq N$}
		\STATE $\hat{x_{i,j}}=\frac{\hat{CCM_{i}}}{10^{N-j-1}}$
		\ENDWHILE
		\RETURN $\hat{X_{i}}$
	\end{algorithmic}
\end{algorithm}

\section{Experiments}
\label{Experiments}

The following experiments were carried out to validate the proposed method in the context of simple questions, where ``simple'' means the core items of a question are explicit and without any ambiguity. This paper selects two simple and typical comparison questions from two courses. Question I is ``What are similarities and differences between array and pointer?''. Question II is ``What are similarities and differences between packet switching and circuit switching?''.

\subsection{Data and Online Learning System}

A developed online learning environment system based on KM is shown in Fig.~\ref{fig_8}, where Fig.~\ref{fig_8a} illustrates ``The C Programming Language'' course, and Fig.~\ref{fig_8b} depicts ``Computer Network Principle'' course. The left-hand side of “The C Programming Language” interface presents a partial KM of the course, while the right-hand side provides the specific explanation of a $ku$ in KM. Likewise, the left-hand side of ``Computer Network Principle'' interface shows related teaching resources (e.g. PowerPoint), while the middle block is the explanation of a $ku$, and the right-hand side is a partial KM. Mappings between teaching resources and knowledge units are associated at knowledge unit level, i.e., by clicking a $ku$, a partial KM related to $ku$ will display on the right-hand side. Moreover, the middle part will give the $ku's$ explanation regarding its source and specific meaning in order to help learners better master the knowledge unit. The learning behaviors of learners can be accurately acquired by a log collection tool running in backend. A learning log contains $id$, $user\_id$, $user\_name$, $question\_id$, $action\_type$, $object\_id$, $action\_object$ and $timestamp$.

The log data of these two comparison questions was collected during two semesters from 173 participants from the department of computer science and technology of Xi'an Jiaotong University, and some students from Xi'an University of Posts and Telecommunications. After preprocessing data, such as removing records that are unrelated to the comparison question learning, including log in, exit, submit and post, we obtained 10,872 effective records. Table \ref{table_1} shows learning records of Question I and Question II, respectively.

\begin{table}[!t]
	\centering
	\begin{threeparttable}
		\caption{Specification of Log Data}
		\label{table_1}
		\begin{tabular}{lcc}
			\hline
			\hline
			Comparision Questions & Number of Subjects & Number of Records \\
			\hline
			Question I & $104$ & $7664$ \\
			Question II & $69$ & $3208$\\
			\hline
			\hline
		\end{tabular}
	\end{threeparttable}
\end{table}

\begin{figure*}[!t]
	\centering
	\subfloat[Visualization of the first strategy process.]{\includegraphics[width=2.3in]{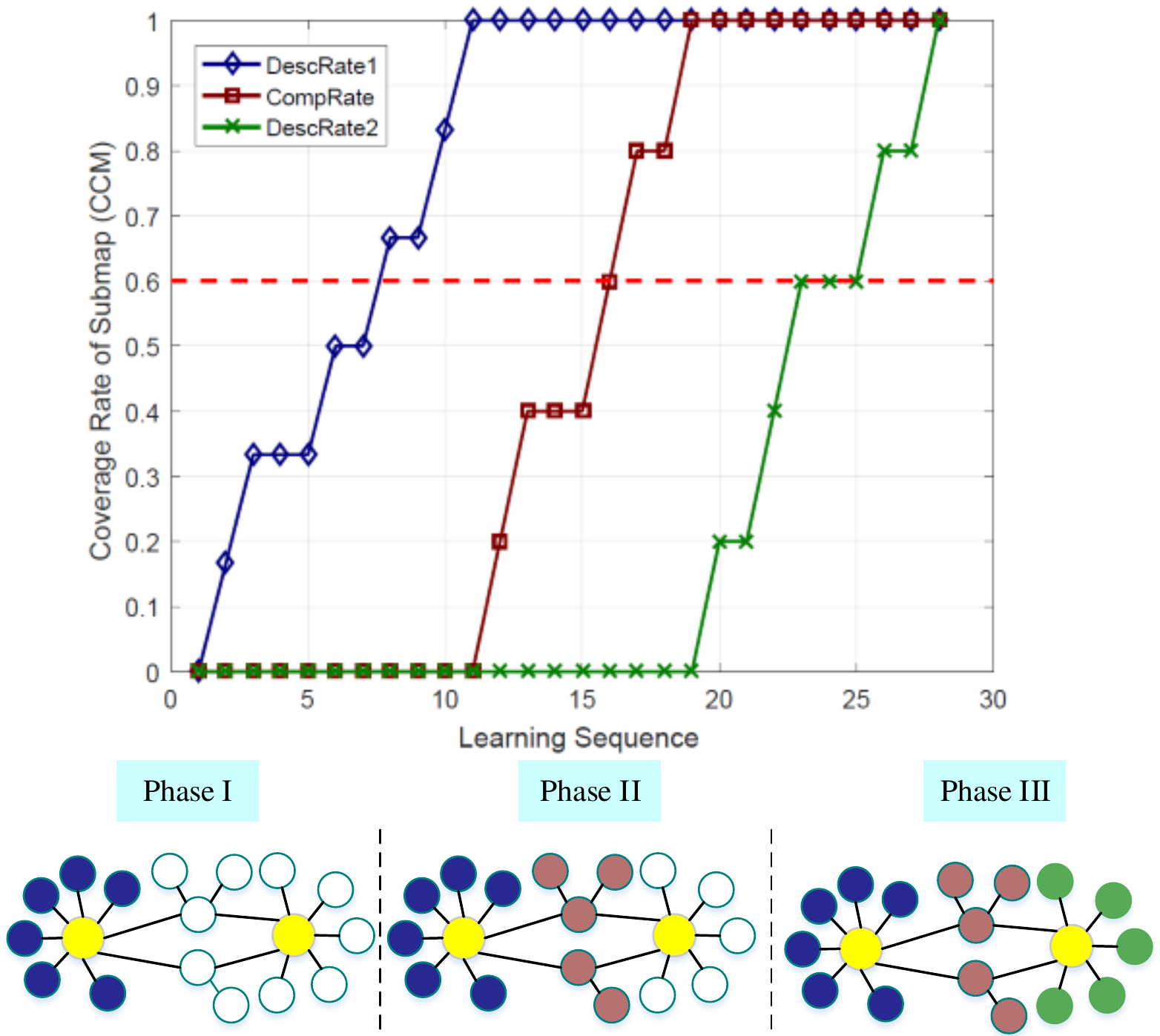}%
		\label{fig_9a}}
	\hfil
	\subfloat[Visualization of the second strategy process.]{\includegraphics[width=2.3in]{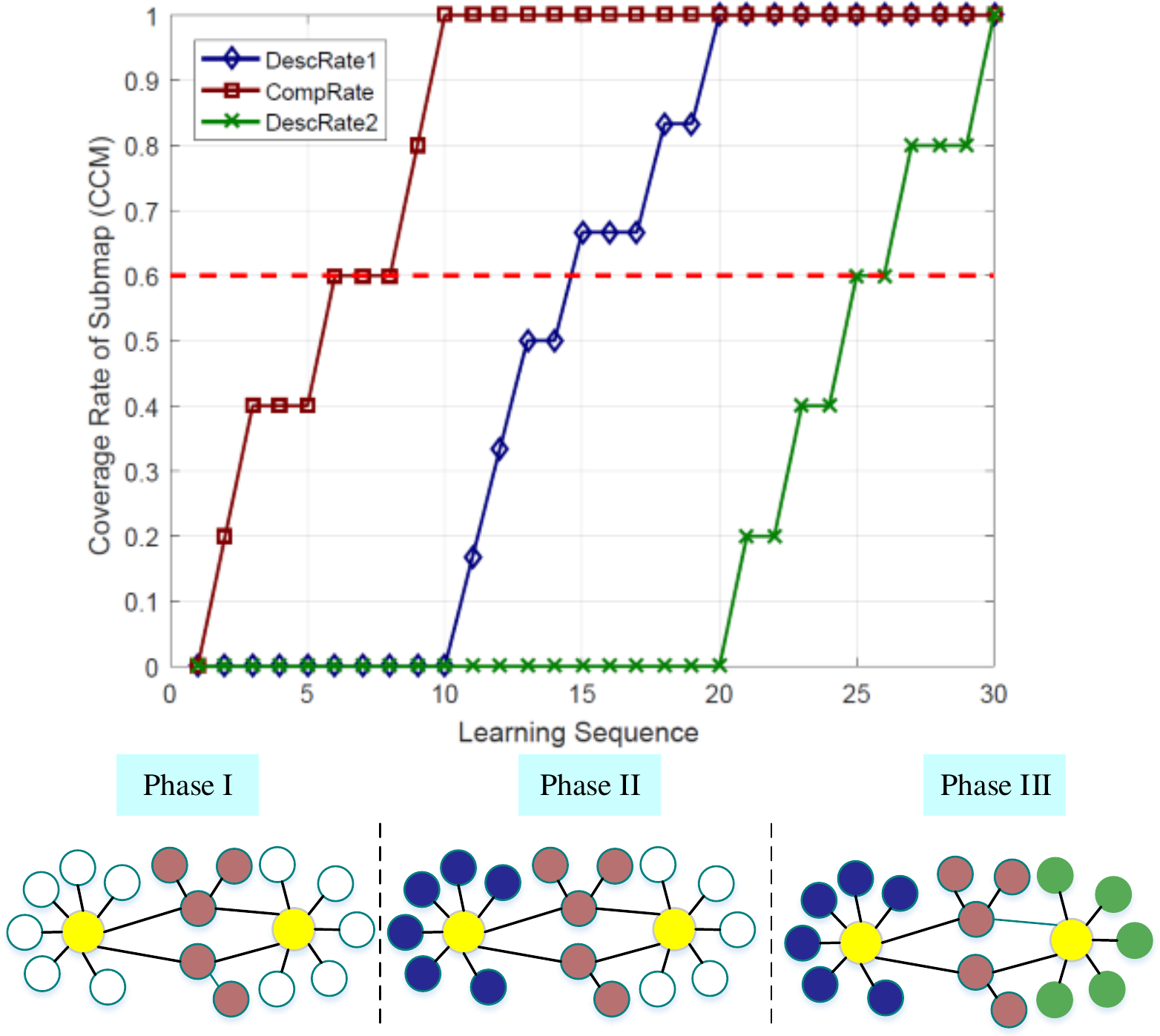}%
		\label{fig_9b}}
	\hfil
	\subfloat[Visualization of the third strategy process.]{\includegraphics[width=2.3in]{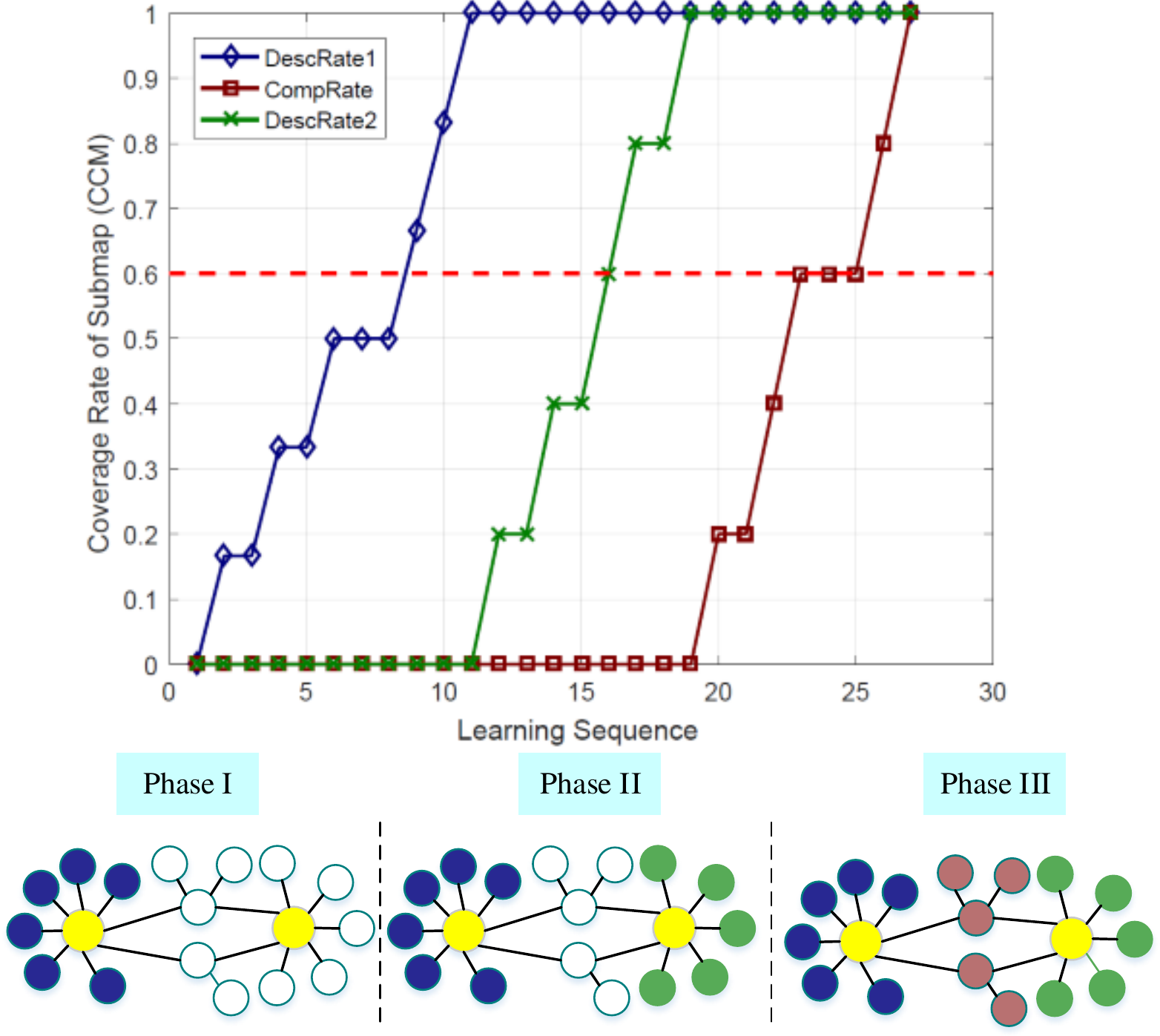}%
		\label{fig_9c}}
	\caption{The Mined three metacognitive strategy patterns.}
	\label{fig_9}
\end{figure*}

\subsection{Evaluation}

After conducting the experiments on 173 subjects, three metacognitive strategy patterns are mined from 10,872 learning records, which are ``Description-Comparison-Description'', ``Comparison-Description-Description'' and ``Description-Description-Comparison''. Additionally, we found that the most frequently occurred submaps are Bubble-Map-like submaps and Double-Bubble-Map-like submaps from participants' learning activity sequences in the task of solving Question I or II.

Three representative cases using cognitive and metacognitive strategy to solve the two questions are visualized in Fig.~\ref{fig_9a}, \ref{fig_9b} and \ref{fig_9c}, respectively. Each figure is comprised of two parts, the top part is a chart consisting of three curves, and the bottom part is a graph-based structure. The horizontal axis of each curve denotes learning event sequence, and the vertical axis denotes knowledge submap coverage rate. Specially, the blue diamond curve represents the change of the descriptive knowledge submap coverage rate of the first $ku$ (e.g., \emph{Array} or \emph{Packet Switching}), $DescRate_{1}$, and the $kus$ of this submap will be tinted with the same dye as soon as being visited; the green x-mark curve represents the change of the descriptive knowledge submap coverage rate of the second $ku$ (e.g., \emph{Pointer} or \emph{Circuit Switching}), $DescRate_{2}$, and the $kus$ of this submap will be tinted in green as well once being visited; the brown square curve shows the change of the connective knowledge submap coverage rate of two $kus$, $CompRate$, similarily, the visited $kus$ of this submap will be colored in brown. What's more, there is a coverage rate threshold line with a value of 0.6, colored in red, which means a valid cognitive strategy is recognized once the coverage rate beyond the threshold.

As shown in Fig.~\ref{fig_9a}, within the interval [1,12], the descriptive knowledge submap coverage rate of the first $ku$, $DescRate_{1}$, changes from 0 to 1, which means the learner finish the visit of all knowledge units describing the first $ku$ during this phase. We can see in [3,5] interval, the knowledge coverage rate stays the same value, which can be ascribed to the subnodes learning ($ku_{4}$ and $ku_{5}$ are the subnodes of $ku_{3}$). However, the descriptive knowledge submap coverage rate of the second $ku$, $DescRate_{2}$, and the connective knowledge submap coverage rate, $CompRate$, remain $0$. This indicates that learners do not visit the $kus$ within the descriptive knowledge submap of the second $ku$ or the connective knowledge submap of two $kus$. Meanwhile, the corresponding operations on these submaps are illustrated in the bottom part of this figure. Seen from the Double Bubble Map that the $kus$ of descriptive knowledge submap of the first $ku$ are visited first and colored in blue, and other $kus$ are not. It is obvious that this learner select learning activities well and the scope of cognitive strategy called description of first $ku$. Hence, this phase is simply deemed ``The First Description Stage''.

Within the interval [13,19], $DescRate_{1}$ and $DescRate_{2}$ keep unchanged, while $CompRate$ changes from 0 to 0.8, which implies learners merely visit the connective knowledge submap of two $kus$. Synchronously, the $kus$ of connective knowledge submap of the first $ku$ and the second $ku$ are visited and colored in red. This phase concentrates on comparison of two $kus$, so it can be called ``Comparison Stage''.

Within the interval [20,24], $DescRate_{1}$ and $CompRate$ remain invariable, while $DescRate_{2}$ changes from 0 to 0.8, concurrently, the $kus$ of descriptive of the second $ku$ are visited and colored in green.Likewise, this phase is ``The Second Description Stage''. Consequently, the question-solving process is accomplished. 

The above three stages in Fig.~\ref{fig_9a} demonstrate that learners control well in the selection of learning activities and the scope of cognitive strategy, meanwhile, the cognition control sequence is $<$descriptive of first $ku$, connective of two $kus$, descriptive of second $ku$$>$. Furthermore, this kind of metacognition strategy pattern is named ``\emph{Description-Comparison-Description}''. Similarly, from Fig.~\ref{fig_9b} and \ref{fig_9c}, we can also conclude  the second and third kind of metacognition strategy pattern can be called ``\emph{Comparison-Description-Description}'' and ``\emph{Description-Description-Comparison}'', respectively.

The experimental results of the three metacognition strategy patterns are shown in Table \ref{table_2}. We can see that the percentage of students who adopted the three metacognition strategy patterns are all close to 31\%, meanwhile, the total sum of these being 93.9\%, which means the majority of students tend to use metacognition strategies to solve the comparative questions and successfully.  Learners who adopt ``\emph{Description-Comparison-Description}'' pattern prefer a sequential manner, and solve the questions step-by-step, however, learners who use the other two patterns may be in a saltaory thinking, and solve questions in general. Therefore, the mining patterns are interpretable and useful in the context of comparative question solving.  At the same time, with regard to the experiment results, the $CCM$ of each description/comparison knowledge submap that students visited is 100\%. For knowing the reason of this, we revisited the students, and they said that they tried their best to find all knowledge units possibly related to the questions during the experiment. So, we believe that this is the reason to have a high value to each $CCM$.

\begin{table}[!t]
	\centering
	\begin{threeparttable}
		\caption{Use of Metacognition Strategy Patterns}
		\label{table_2}
		\begin{tabular}{lcc}
			\hline
			\hline
			Metacognition Strategy Pattern & Percentage & Sum \\
			\hline
			Description-Comparison-Description & $31.3\%$ &  \\
			Comparison-Description-Description & $31.0\%$ & $93.9\%$ \\
			Description-Description-Comparison & $31.6\%$ &  \\
			\hline
			\hline
		\end{tabular}
	\end{threeparttable}
\end{table}

\section{Conclusion and Future Work}

This study focuses on how learners solve questions by adopting their cognitive and metacognitive strategies, which follows an idea of reverse engineering. We propose a novel method that combines Thinking Maps with Knowledge Maps to detect and model the cognitive and metacognitive strategies from the learning logs gathered from a computer-based e-learning system. From the experimental results, the idea of raising the level of abstraction is quite important, which verifies that cognitive strategies can be described with a domain specific knowledge map, and metacognition strategies which are complex constructs that are not directly observable, can be obtained from a cognition control sequence that consist of cognitive strategies. This shows that mapping between cognitive strategies and procedural cognitive knowledge (metacognition) can be automated to provide their meaningful interactions and supports. Meanwhile, the cognitive strategies and three metacognitive strategy patterns can be represented in a graph structure. Consequently, this graph data-driven method is proved to be effective, as it provides a way to visualize the thinking process of solving questions.

It should be noted that this research has some limitations. Firstly, we do not consider the differences in student's background, knowledge and their general cognitive ability. Many researches indicated that the difference has played an important role in the use of cognitive strategies and metacognition strategies. Secondly, we only discuss use of cognitive and metacognitive strategy within the scope of comparative question-solving in this paper. Thirdly, we conduct experiments on simple questions, not considering the ambiguity of questions and reduce its complexity. In future work, we will expand not only the number of participants including various topics for comparison, but also the question types. Additional, we will take natural language related contents into account, to improve our methods and algorithms by natural language processing techniques.

\section*{Acknowledgments}
This work was supported by the National Key Research and Development Program of China under Grant No. 2016YFB1000903; the National Science Foundation of China under Grant No. 61877048; Innovative Research Group of the National Natural Science Foundation of China under Grant No. 61721002; Innovation Research Team of Ministry of Education (IRT\_17R86); project of China Knowledge Centre for Engineering Science and Technology. 

%\section*{References}
\bibliographystyle{IEEEtran}
\bibliography{IEEEabrv,ref}

\begin{thebibliography}{10}
\providecommand{\url}[1]{#1}
\csname url@samestyle\endcsname
\providecommand{\newblock}{\relax}
\providecommand{\bibinfo}[2]{#2}
\providecommand{\BIBentrySTDinterwordspacing}{\spaceskip=0pt\relax}
\providecommand{\BIBentryALTinterwordstretchfactor}{4}
\providecommand{\BIBentryALTinterwordspacing}{\spaceskip=\fontdimen2\font plus
\BIBentryALTinterwordstretchfactor\fontdimen3\font minus
  \fontdimen4\font\relax}
\providecommand{\BIBforeignlanguage}[2]{{%
\expandafter\ifx\csname l@#1\endcsname\relax
\typeout{** WARNING: IEEEtran.bst: No hyphenation pattern has been}%
\typeout{** loaded for the language `#1'. Using the pattern for}%
\typeout{** the default language instead.}%
\else
\language=\csname l@#1\endcsname
\fi
#2}}
\providecommand{\BIBdecl}{\relax}
\BIBdecl

\bibitem{kim2016applying}
J.~H. Kim, L.~Rothrock, and A.~Tharanathan, ``Applying fuzzy linear regression
  to understand metacognitive judgments in a human-in-the-loop simulation
  environment,'' \emph{IEEE Transactions on Human-Machine Systems}, vol.~46,
  no.~3, pp. 360--369, 2016.

\bibitem{2le2018cognitive}
N.-T. Le and L.~Wartschinski, ``A cognitive assistant for improving human
  reasoning skills,'' \emph{International Journal of Human-Computer Studies},
  2018.

\bibitem{5winne2010bootstrapping}
P.~H. Winne, ``Bootstrapping learner's self-regulated learning,''
  \emph{Psychological Test and Assessment Modeling}, vol.~52, no.~4, p. 472,
  2010.

\bibitem{41Flavell1979Metacognition}
J.~H. Flavell, ``Metacognition and cognitive monitoring: A new area of
  cognitive–developmental inquiry.'' \emph{American Psychologist}, vol.~34,
  no.~10, pp. 906--11, 1979.

\bibitem{1Simon2004Learning}
S.~Cassidy, ``Learning styles: An overview of theories, models, and measures,''
  \emph{Educational Psychology}, vol.~24, no.~4, pp. 419--444, 2004.

\bibitem{3Metcalfe2013Metacognition}
Metcalfe, Janet, Finn, and Bridgid, ``Metacognition and control of study choice
  in children,'' \emph{Metacognition \& Learning}, vol.~8, no.~1, pp. 19--46,
  2013.

\bibitem{13pintrich1993reliability}
P.~R. Pintrich, D.~A. Smith, T.~Garcia, and W.~J. McKeachie, ``Reliability and
  predictive validity of the motivated strategies for learning questionnaire
  (mslq),'' \emph{Educational and psychological measurement}, vol.~53, no.~3,
  pp. 801--813, 1993.

\bibitem{14weinstein1987lassi}
C.~E. Weinstein, A.~C. Schulte, and A.~W. Hoy, \emph{LASSI: Learning and study
  strategies inventory}.\hskip 1em plus 0.5em minus 0.4em\relax H \& H
  Publishing Company, 1987.

\bibitem{17hadwin2007examining}
A.~F. Hadwin, J.~C. Nesbit, D.~Jamieson-Noel, J.~Code, and P.~H. Winne,
  ``Examining trace data to explore self-regulated learning,''
  \emph{Metacognition and Learning}, vol.~2, no. 2-3, pp. 107--124, 2007.

\bibitem{19perry2006learning}
N.~E. Perry and P.~H. Winne, ``Learning from learning kits: gstudy traces of
  students’ self-regulated engagements with computerized content,''
  \emph{Educational Psychology Review}, vol.~18, no.~3, pp. 211--228, 2006.

\bibitem{21kinnebrew2011comparative}
J.~Kinnebrew and G.~Biswas, ``Comparative action sequence analysis with hidden
  markov models and sequence mining,'' in \emph{Proceedings of the Knowledge
  Discovery in Educational Data Workshop at the 17th ACM SIGKDD Conference on
  Knowledge Discovery and Data Mining (KDD 2011). San Diego, CA}, 2011.

\bibitem{24biswas2013analyzing}
G.~Biswas, J.~Kinnebrew, and J.~Segedy, ``Analyzing students' metacognitive
  strategies in open-ended learning environments,'' in \emph{Proceedings of the
  Annual Meeting of the Cognitive Science Society}, vol.~35, no.~35, 2013.

\bibitem{25kinnebrew2014analyzing}
J.~S. Kinnebrew, J.~R. Segedy, and G.~Biswas, ``Analyzing the temporal
  evolution of students’ behaviors in open-ended learning environments,''
  \emph{Metacognition and learning}, vol.~9, no.~2, pp. 187--215, 2014.

\bibitem{26segedy2015using}
J.~R. Segedy, J.~S. Kinnebrew, and G.~Biswas, ``Using coherence analysis to
  characterize self-regulated learning behaviours in open-ended learning
  environments,'' \emph{Journal of Learning Analytics}, vol.~2, no.~1, pp.
  13--48, 2015.

\bibitem{27perera2009clustering}
D.~Perera, J.~Kay, I.~Koprinska, K.~Yacef, and O.~R. Za{\"\i}ane, ``Clustering
  and sequential pattern mining of online collaborative learning data,''
  \emph{IEEE Transactions on Knowledge and Data Engineering}, vol.~21, no.~6,
  pp. 759--772, 2009.

\bibitem{28gavsevic2017detecting}
D.~Ga{\v{s}}evic, J.~Jovanovic, A.~Pardo, and S.~Dawson, ``Detecting learning
  strategies with analytics: Links with self-reported measures and academic
  performance.'' \emph{Journal of Learning Analytics}, vol.~4, no.~2, pp.
  113--128, 2017.

\bibitem{29fern2010mining}
X.~Fern, C.~Komireddy, V.~Grigoreanu, and M.~Burnett, ``Mining problem-solving
  strategies from hci data,'' \emph{ACM Transactions on Computer-Human
  Interaction (TOCHI)}, vol.~17, no.~1, p.~3, 2010.

\bibitem{30crockett2017predicting}
K.~Crockett, A.~Latham, and N.~Whitton, ``On predicting learning styles in
  conversational intelligent tutoring systems using fuzzy decision trees,''
  \emph{International Journal of Human-Computer Studies}, vol.~97, pp. 98--115,
  2017.

\bibitem{29tian2007personalized}
F.~Tian, Q.~Zheng, Z.~Gong, J.~Du, and R.~Li, ``Personalized learning
  strategies in an intelligent e-learning environment,'' in \emph{Computer
  Supported Cooperative Work in Design, 2007. CSCWD 2007. 11th International
  Conference on}.\hskip 1em plus 0.5em minus 0.4em\relax IEEE, 2007, pp.
  973--978.

\bibitem{9josyula2009modeling}
D.~P. Josyula, H.~Vadali, B.~J. Donahue, and F.~C. Hughes, ``Modeling
  metacognition for learning in artificial systems,'' in \emph{Nature \&
  Biologically Inspired Computing, 2009. NaBIC 2009. World Congress on}.\hskip
  1em plus 0.5em minus 0.4em\relax IEEE, 2009, pp. 1419--1424.

\bibitem{23segedy2011modeling}
J.~R. Segedy, J.~S. Kinnebrew, and G.~Biswas, ``Modeling learner's cognitive
  and metacognitive strategies in an open-ended learning environment.'' in
  \emph{AAAI Fall Symposium: Advances in Cognitive Systems}, 2011.

\bibitem{22kinnebrew2013contextualized}
J.~S. Kinnebrew, K.~M. Loretz, and G.~Biswas, ``A contextualized, differential
  sequence mining method to derive students' learning behavior patterns,''
  \emph{JEDM| Journal of Educational Data Mining}, vol.~5, no.~1, pp. 190--219,
  2013.

\bibitem{10biswas2010measuring}
G.~Biswas, H.~Jeong, J.~S. Kinnebrew, B.~Sulcer, and R.~ROSCOE, ``Measuring
  self-regulated learning skills through social interactions in a teachable
  agent environment,'' \emph{Research and Practice in Technology Enhanced
  Learning}, vol.~5, no.~02, pp. 123--152, 2010.

\bibitem{20kinnebrew2013investigating}
J.~S. Kinnebrew, G.~Biswas, B.~Sulcer, and R.~S. Taylor, ``Investigating
  self-regulated learning in teachable agent environments,'' in
  \emph{International handbook of metacognition and learning
  technologies}.\hskip 1em plus 0.5em minus 0.4em\relax Springer, 2013, pp.
  451--470.

\bibitem{33Hyerle2014Thinking}
D.~Hyerle, \emph{Thinking Maps®: A Visual Language for Learning}.\hskip 1em
  plus 0.5em minus 0.4em\relax Springer London, 2008.

\bibitem{34Liu2012Topological}
J.~Liu, J.~Wang, Q.~Zheng, W.~Zhang, and L.~Jiang, ``Topological analysis of
  knowledge maps,'' \emph{Knowledge-Based Systems}, vol.~36, no.~6, pp.
  260--267, 2012.

\bibitem{35O2002Knowledge}
A.~M. O'Donnell, D.~F. Dansereau, and R.~H. Hall, ``Knowledge maps as scaffolds
  for cognitive processing,'' \emph{Educational Psychology Review}, vol.~14,
  no.~1, pp. 71--86, 2002.

\bibitem{36Zheng2011Yotta}
Q.~Zheng, Y.~Qian, and J.~Liu, ``Yotta: A knowledge map centric e-learning
  system,'' in \emph{IEEE International Conference on E-Business Engineering},
  2011, pp. 42--49.

\bibitem{37Zeng2014A}
B.~Zeng, F.~Tian, B.~Feng, Q.~Zheng, F.~Wu, and Y.~Fu, ``A study on algorithm
  for mining subgraph in thinking maps,'' in \emph{IEEE International
  Conference on High PERFORMANCE Computing and Communications \& 2013 IEEE
  International Conference on Embedded and Ubiquitous Computing}, 2014, pp.
  2368--2372.

\bibitem{38Huang2014Generating}
X.~J. Huang, C.~Zhang, and Q.~H. Zheng, ``Generating personalized navigation
  learning path based on knowledge map,'' \emph{International Journal of
  Technology \& Educational Marketing}, vol.~4, no.~2, pp. 1--17, 2014.

\bibitem{39ocford}
O.~dictionary, ``cognition - definition of cognition in english from the oxford
  dictionary,'' \url{www.oxforddictionaries.com}, 2016.

\bibitem{40Pressley1995Cognitive}
M.~Pressley, ``Cognitive strategy instruction that really improves children's
  academic performance,'' \emph{Academic Achievement}, p. 266, 1995.

\bibitem{42Cross1988Developmental}
D.~R. Cross, ``Developmental and instructional analyses of children's
  metacognition and reading comprehension.'' \emph{Journal of Educational
  Psychology}, vol.~80, no.~2, pp. 131--142, 1988.

\bibitem{43Paris1990Promoting}
S.~G. Paris and P.~Winograd, ``Promoting metacognition and motivation of
  exceptional children.'' \emph{Remedial \& Special Education}, vol.~11, no.~6,
  pp. 7--15, 1990.

\bibitem{44Schraw1995Metacognitive}
G.~Schraw and D.~Moshman, ``Metacognitive theories,'' \emph{Educational
  Psychology Review}, vol.~7, no.~4, pp. 351--371, 1995.

\bibitem{45Schraw2006Promoting}
G.~Schraw, K.~J. Crippen, and K.~Hartley, ``Promoting self-regulation in
  science education: Metacognition as part of a broader perspective on
  learning,'' \emph{Research in Science Education}, vol.~36, no. 1-2, pp.
  111--139, 2006.

\bibitem{46Deanna2004Metacognition}
D.~Kuhn and J.~David~Dean, ``Metacognition: A bridge between cognitive
  psychology and educational practice.'' \emph{Theory Into Practice}, vol.~43,
  no.~4, pp. 268--273, 2004.

\bibitem{47Montague1992The}
M.~Montague, ``The effects of cognitive and metacognitive strategy instruction
  on the mathematical problem solving of middle school students with learning
  disabilities.'' \emph{Journal of Learning Disabilities}, vol.~25, no.~4, p.
  230, 1992.

\bibitem{7zimmerman1986development}
B.~J. Zimmerman and M.~M. Pons, ``Development of a structured interview for
  assessing student use of self-regulated learning strategies,'' \emph{American
  educational research journal}, vol.~23, no.~4, pp. 614--628, 1986.

\bibitem{48J7368935}
J.~S. Kinnebrew, J.~R. Segedy, and G.~Biswas, ``Integrating model-driven and
  data-driven techniques for analyzing learning behaviors in open-ended
  learning environments,'' \emph{IEEE Transactions on Learning Technologies},
  vol.~10, no.~2, pp. 140--153, April 2017.

\bibitem{2Leelawong2008Designing}
K.~Leelawong, ``Designing learning by teaching systems : The betty's brain
  system,'' \emph{Int.j.artif.intell.in Education}, 2008.

\bibitem{49Riding2011The}
R.~J. Riding and G.~Douglas, ``The effect of cognitive style and mode of
  presentation on learning performance.'' \emph{British Journal of Educational
  Psychology}, vol.~63, no.~2, pp. 297--307, 1993.

\bibitem{50Rezaei2004Evaluation}
A.~R. Rezaei and L.~Katz, ``Evaluation of the reliability and validity of the
  cognitive styles analysis,'' \emph{Personality \& Individual Differences},
  vol.~36, no.~6, pp. 1317--1327, 2004.

\bibitem{51Hmelo2004Problem}
C.~E. Hmelo-Silver, ``Problem-based learning: What and how do students learn?''
  \emph{Educational Psychology Review}, vol.~16, no.~3, pp. 235--266, 2004.

\bibitem{52bailey2005introduction}
T.~Bailey, ``An introduction to the c programming language and software
  design,'' 2005.

\bibitem{53Srikant1996Mining}
R.~Srikant and R.~Agrawal, ``Mining sequential patterns: Generalizations and
  performance improvements,'' in \emph{International Conference on Extending
  Database Technology}, 1996, pp. 1--17.

\end{thebibliography}

\ifCLASSOPTIONcaptionsoff
  \newpage
\fi



%\bibliographystyle{IEEEtran}

\begin{IEEEbiography}[{\includegraphics[width=1in,height=1.25in,clip,keepaspectratio]{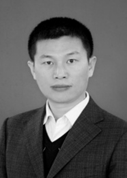}}]{Feng Tian}
	was born in Xi'an, ShaanXi, China in 1972. He received the B.S. degree in industrial automation and the M.S. degree in computer science and technology from the Xi'an University of Architecture and Technology, Xi'an, China, in 1995 and 2000, respectively, and the Ph.D. degree in control theory and application from Xi'an Jiaotong University, Xi'an, in 2003.
	
	He has been with Xi'an Jiaotong University since 2004, where he is currently with National Engineering Lab of Big Data Analytics and also with the Systems Engineering Institute, as a Professor. He is a member of the Satellite-Terrestrial Network Technology R\&D Key Laboratory, Shaanxi Province. His research interests include big data analytics, learning analytics, system modelling and analysis, and cloud computing. He is a member of the IEEE.
\end{IEEEbiography}

\begin{IEEEbiography}[{\includegraphics[width=1in,height=1.25in,clip,keepaspectratio]{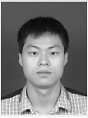}}]{Jia Yue}
	received the B.S degree in automation from Xidian University in 2014. He is currently pursing in the PhD degree in Satellite-Terrestrial Network Technology R\&D Key Laboratory at Xi'an Jiaotong University, focusing on data mining and multimodal learning.
\end{IEEEbiography}

\begin{IEEEbiography}[{\includegraphics[width=1in,height=1.25in,clip,keepaspectratio]{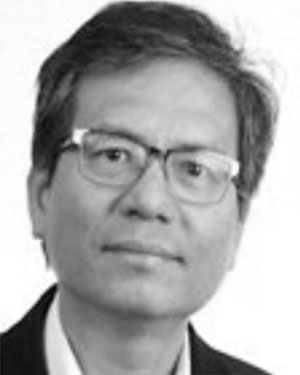}}]{Kuo-Ming Chao}
	is a professor of computing at Coventry University, UK. His research interests include the areas of cloud computing and big data etc as well as their applications. He has over 150 refereed publications. He is a member of the IEEE.
\end{IEEEbiography}

\begin{IEEEbiography}[{\includegraphics[width=1in,height=1.25in,clip,keepaspectratio]{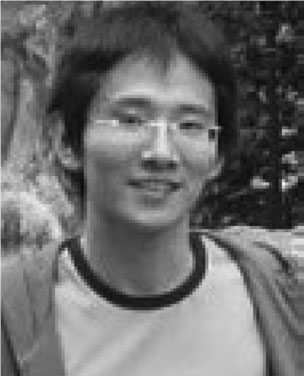}}]{Buyue Qian}
	received the BS degree in information engineering from Xi'an Jiaotong University, in 2007, and the Master of Science degree from Columbia University, in 2009. He received the PhD degree from the Department of Computer Science, University of California at Davis, in 2013. He is currently a research scientist at IBM T.J. Watson Research. He received the Yahoo! Research Award, the IBM Eminence and Excellence Award, and the SIAM Data Mining2013 Best Research Paper Runner Up Award.
\end{IEEEbiography}

\begin{IEEEbiography}[{\includegraphics[width=1in,height=1.25in,clip,keepaspectratio]{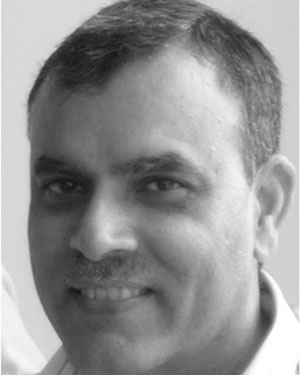}}]{Nazaraf Shah }
	is a senior lecturer at Coventry University, UK. He interested in intelligent agent and cloud computing. He has over 50 publications in various international conferences and journals.
\end{IEEEbiography}

\begin{IEEEbiography}[{\includegraphics[width=1in,height=1.25in,clip,keepaspectratio]{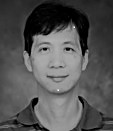}}]{Longzhuang Li}
	 is a professor at Texas A\&M University-Corpus Christi. His research focuses on data integration and data mining and has been supported by National Science Foundation and Air Force Research Lab.
\end{IEEEbiography}

\begin{IEEEbiography}[{\includegraphics[width=1in,height=1.25in,clip,keepaspectratio]{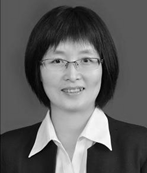}}]{Haiping Zhu}
	received the B.S. degree in electronic science from Northwest University in 1996, the M.S. degree in information and communication engineer from Chinese Academy of Sciences in 1999, and the Ph.D. degree in computer science from Xi'an Jiaotong University in 2009.
	
	She is currently a Lecturer and graduate supervisor with the Department of Computer Science and Technology, Xi'an Jiaotong University. Her research interests include educational data mining, learning analytics and personalized recommendation.
	
\end{IEEEbiography}

\begin{IEEEbiography}[{\includegraphics[width=1in,height=1.25in,clip,keepaspectratio]{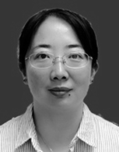}}]{Yan Chen}
	received the B.S. degree in computer science and technology from Xi'an Jiaotong University, in 1993, the M.S. degree in computer science and technology from Xi'an Jiaotong University, in 1996, and the Ph.D. degree in computer science from Xi'an Jiaotong University, China in 2005.
	
	She is currently an Associate Professor and graduate supervisor with the Department of Computer Science and Technology, Xi'an Jiaotong University. Her research interests include educational data mining and learning analytics.
\end{IEEEbiography}

\begin{IEEEbiography}[{\includegraphics[width=1in,height=1.25in,clip,keepaspectratio]{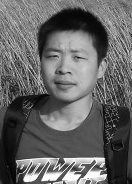}}]{Bing Zeng}
	received the B.S degree in mathematics from Chang'an University and M.S degree from Xi'an Jiaotong University, in 2012 and 2015 respectively.
\end{IEEEbiography}

\begin{IEEEbiography}[{\includegraphics[width=1in,height=1.25in,clip,keepaspectratio]{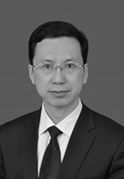}}]{Qinghua Zheng}
	received the B.S. degree in computer software, the M.S. degree in computer organization and architecture, and the Ph.D. degree in system engineering from Xi'an Jiaotong University, China, in 1990, 1993, and 1997,respectively.
	
	He did post-doctoral research at Harvard University in 2002 and was a Visiting Professor of Research with the Hong Kong University from 2004 to 2005. 
	Dr. Zheng received the First Prize for National Teaching Achievement, State Education Ministry in 2005 and the First Prize for Scientific and Technological Development of Shanghai City and Shaanxi Province, in 2004 and 2003, respectively.
\end{IEEEbiography}

\end{document}